\documentclass[11pt]{article}
\usepackage{a4wide}
\usepackage{amsmath,amssymb}
\usepackage{simplewick}

\usepackage{mathrsfs}

\usepackage{slashed}
\usepackage{graphicx}
\usepackage{color}
\usepackage{mathtools}
\usepackage{comment}
\usepackage{cite}
\usepackage[bookmarks=true,bookmarksnumbered=true,setpagesize=false]{hyperref}

\newcommand{\eref}[1]{(\ref{#1})}

\usepackage{enumerate}
\newcommand{\al}[1]{\begin{align}#1\end{align}}

\newcommand{\ov}{\over}

\newcommand{\tx}{\text}

\newcommand{\paren}[1]{\left(#1\right)}
\newcommand{\pn}[1]{\left(#1\right)}

\newcommand{\sqbr}[1]{\left[#1\right]}
\newcommand{\ab}[1]{\left|#1\right|}

\newcommand{\br}[1]{\left\{#1\right\}}
\newcommand{\fn}[1]{\!\left(#1\right)}

\newcommand{\Pn}[1]{\bigl(#1\bigr)}

\newcommand{\df}{\text{d}}

\newcommand{\I}{\text{I}}
\newcommand{\R}{\text{R}}

\newcommand{\p}{\partial}

\newcommand{\wt}{\widetilde}

\newcommand{\wh}{\widehat}

\usepackage{braket}

\newcommand{\vep}{\varepsilon}

\newcommand{\LOeq}{\stackrel{\mathclap{\mbox{\scriptsize LO}}}{=}}

\usepackage[normalem]{ulem}

\allowdisplaybreaks[2]

\begin{document}

%

\title{
\vspace{-2cm}
\vbox{
\baselineskip 14pt
\hfill \hbox{\normalsize RIKEN-iTHEMS-Report-23
}} 
\vskip 1cm
\bf \Large
Gradient-flowed order parameter\\
for spontaneous gauge symmetry breaking
\vskip 0.5cm
}
\author{
Kengo Kikuchi,\thanks{E-mail: \tt kengo@yukawa.kyoto-u.ac.jp} \mbox{}
Kenji Nishiwaki,\thanks{E-mail: \tt kenji.nishiwaki@snu.edu.in} \mbox{}
and
Kin-ya Oda\thanks{E-mail: \tt odakin@lab.twcu.ac.jp}
\bigskip\\
}

\date{\today}
\maketitle

\begin{center}
\it
$^*$ RIKEN iTHEMS, Wako, Saitama 351-0198, Japan\\
$^\dagger$ Department of Physics, Shiv Nadar Institution of Eminence,\\ Gautam Buddha Nagar, 201314, India\\
$^\ddag$ Department of Mathematics, Tokyo Woman's Christian University, Tokyo 167-8585, Japan\\
\end{center}
\rm
\begin{abstract}\noindent
The gauge-invariant two-point function of the Higgs field at the same spacetime point can make a natural gauge-invariant order parameter for spontaneous gauge symmetry breaking. However, this composite operator is ultraviolet divergent and is not well defined. We propose using a gradient flow to cure the divergence from putting the fields at the same spacetime point. As a first step, we compute it for the Abelian Higgs model with a positive mass squared at the one-loop order in the continuum theory using the saddle-point method to estimate the finite part. The order parameter consistently goes to zero in the infrared limit of the infinite flow time.
\end{abstract}

\newpage 
\tableofcontents

\newpage

\section{Introduction}
A non-perturbative definition of gauge theory has so far been given only on a lattice.
A gauge symmetry cannot be spontaneously broken by the Higgs mechanism when one path integrates over the whole gauge configurations on the lattice~\cite{Elitzur:1975im}.
In the continuum perturbative calculation, the vacuum expectation value (VEV) of the Higgs field, $\bigl\langle\wh\Phi\fn{x}\bigr\rangle$, is regarded as the order parameter of the spontaneous symmetry breaking (SSB).
The crucial argument for the impossibility of the SSB in the non-perturbative lattice gauge theory is that $\wh\Phi\fn{x}$ is not a gauge-invariant operator, and hence its expectation value $\bigl\langle\wh\Phi\fn{x}\bigr\rangle=\int{{\cal D}\Phi}\,\Phi\fn{x}e^{-S[\Phi]}$ becomes zero, due to the Elitzur theorem when one path integrates over all the gauge configurations without a gauge fixing~\cite{Elitzur:1975im,Frohlich:1980gj,Frohlich:1981yi}.
In lattice gauge theory, the SSB can be measured, for example, by the two-point correlation function of the Higgs field at two different points, which is not gauge-invariant, by limiting the path integral to a particular fixed gauge slice; see, e.g., Ref.~\cite{Fradkin:1978dv} and also Ref.~\cite{Maas:2017wzi} for a recent review.
It is desirable to have a gauge-invariant order parameter that does not require gauge fixing.

A natural gauge-invariant operator is $\wh\Phi^\dagger\fn{x}\wh\Phi\fn{x}$.
The problem is that this operator suffers from ultraviolet (UV) divergences by placing two operators $\wh\Phi^\dagger\fn{x}$ and $\wh\Phi\fn{x}$ at the same spacetime point $x$.
One might then turn to a two-point function inserted with a path-ordered Wilson line $\wh\Phi^\dagger\fn{x}\tx P\exp\fn{ig\int_y^x\df z^\mu\wh A_\mu\fn{z}}\wh\Phi\fn{y}$ in the infrared (IR) limit $\ab{x-y}\to\infty$.
This operator contains both the Higgs VEV and the Wilson line contribution. Therefore, one must examine the scaling behavior in the limit $\ab{x-y}\to\infty$ in order to distinguish the symmetric, Higgs, and confinement phases.\footnote{Here, the confinement phase denotes when the Wilson line acquires a nontrivial (non-unity) expectation value; the Higgs phase denotes when only the Higgs acquires an expectation value; the symmetric phase denotes when both the expectation values of the Higgs and Wilson line vanish.} It would be worthwhile if one can provide a regularized composite operator for $\wh\Phi^\dagger\fn{x}\wh\Phi\fn{x}$.\footnote{
{Other discussions on gauge dependence in phase structures are found, e.g., in~\cite{Buchmuller:1994vy,Hu:1996qa,DiLuzio:2014bua,Dittmaier:2022maf}.}
}

Gradient flow is a powerful tool to cure UV divergences by smearing the field and going into the extra flow time direction~\cite{Narayanan:2006rf,Luscher:2009eq,Luscher:2010iy,Luscher:2011bx}.
In particular, one can completely remove the UV divergences from placing the gauge fields at the same spacetime point in the correlation functions~\cite{Luscher:2010iy,Luscher:2011bx}.
The same argument holds for the quark fields except for the requirement for an extra field (wave function) renormalization~\cite{Luscher:2013cpa}.
(Various other aspects have been explored in the context of the gradient flow, such as
the non-perturbative renormalization group~\cite{Sint:2014pip,Makino:2018rys,Abe:2018zdc,Carosso:2018bmz,Sonoda:2019ibh,Sonoda:2020vut,Miyakawa:2021hcx,Miyakawa:2021wus,Abe:2022smm,Sonoda:2022fmk,Miyakawa:2022qbz,Hasenfratz:2022wll},
holographic theories~\cite{Aoki:2015dla,Aoki:2016ohw,Aoki:2017bru,Aoki:2017uce,Aoki:2018dmc,Aoki:2019bfb,Aoki:2022lye},
the $O\fn{N}$ nonlinear sigma model and its large-$N$ expansion~\cite{Makino:2014sta,Makino:2014cxa,Aoki:2014dxa,Aoki:2016env},
supersymmetric gradient flow~\cite{Nakazawa:2004ac,Nakazawa:2003tz,Kikuchi:2014rla,Hieda:2017sqq,Aoki:2017iwi,Kasai:2018koz,Kadoh:2018qwg,Bergner:2019dim,Kadoh:2019glu,Kadoh:2019flv,Kadoh:2023gqa,Kadoh:2023mof},
phenomenological applications~\cite{Chigusa:2019wxb,Sato:2019axv,Ho:2019ads,Hamada:2020rnp,Rizik:2020naq,Suzuki:2020zue,Brambilla:2021egm,Mereghetti:2021nkt,Harlander:2022tgk},
and formal issues in quantum field theory~\cite{Suzuki:2013gza,Makino:2014taa,Fujikawa:2016qis,Hieda:2016xpq,Morikawa:2018fek}.)

In this paper, we propose a gauge-invariant order parameter for the SSB using a gradient flow---the \emph{flowed order parameter}, $\Braket{\wh\Psi^\dagger\fn{t,x}\wh\Psi\fn{s,x}}$, where $\Psi^\dagger\fn{t,x}$ and $\Psi\fn{s,x}$ are the classical solutions to the gradient flow equation toward the flow times $t$ and $s$ with the boundary conditions $\Psi^\dagger\fn{0,x}=\Phi^\dagger\fn{x}$ and $\Psi\fn{0,x}=\Phi\fn{x}$, respectively.
This order parameter naturally cures the abovementioned UV divergence from placing the Higgs fields at the same spacetime point thanks to smearing from the gradient flow.

As a first step in one of the simplest examples, we compute the gauge-invariant flowed order parameter in gauge-fixed perturbation theory in the symmetric phase in the Abelian Higgs model, which is asymptotically non-free.

The organization of this paper is as follows.
In Sect.~\ref{2}, we describe our setup of the model and its gradient flow equation.
In Sect.~\ref{3}, we show the UV finiteness of the gauge-boson two-point function at the one-loop level.
In Sect.~\ref{4}, we consider the Higgs two-point function as a flowed order parameter and analyze the behavior of the finite part at the large flow time limit.
Section~\ref{5} provides a summary and discussions.
In Appendix~\ref{sec:Detail-HTPF}, we provide details on the calculations of the Higgs two-point function at the one-loop level.

\section{Gradient flow of Abelian Higgs model}\label{2}
We first present the setup for the analysis in this paper and then show the gradient flow equations and their solutions.
\subsection{Setup}
We study the Abelian Higgs model with a $U(1)$ Higgs field $\Phi$ and a gauge field $A_\mu$ in the $d$-dimensional Euclidean spacetime, with the metric $\delta_{\mu\nu} = \delta^{\mu\nu} = \paren{\I_d}_{\mu\nu} = \paren{\I_d}^{\mu\nu}$,
\al{
S 
	&:= 
		S_A + S_\Phi, 		\label{eq:bulk-action}\\
S_A
	&:=
		\frac{1}{g_0^2} \int \df^d x 
		\br{ \frac{1}{4} F_{\mu\nu} F_{\mu\nu} + \frac{1}{2\xi_0} \paren{ \p_\mu A_\mu }^2 }, \\
S_\Phi
	&:=
		\frac{1}{g_0^2} \int \df^d x \br{ \paren{D_\mu \Phi}^\dagger D_\mu \Phi + m_0^2 \paren{\Phi^\dagger \Phi}
		+ \frac{\lambda_0}{g_0^2} \paren{\Phi^\dagger \Phi}^2 },
		\label{Phi action}
}
where $\I_d$ is the $d$-dimensional identity matrix;
$\Phi$ is a complex scalar field with a unit charge;
$g_0$, $m_0$, $\xi_0$, and $\lambda_0$ are the bare gauge coupling, the bare mass, the bare gauge-fixing parameter, and the bare self-coupling of $\Phi$, respectively;
$F_{\mu\nu}=\p_\mu A_\nu - \p_\nu A_\mu$ is the Abelian field strength; and
the covariant derivative $D_\mu$ takes the form
\al{
D_\mu := \p_\mu - i A_\mu.
}
We have employed the non-canonically normalized fields $\Phi=g_0\Phi^\tx{can}$ and $A_\mu=g_0A^\tx{can}_\mu$, where the superscript ``can'' denotes a canonically normalized field, such that the counting of the number of loops coincides with that of $g_0$ in the gradient flow, as we will see.
Here and hereafter, we write bare fields such as $\Phi$ and $A_\mu$ without any subscript while writing renormalized fields with subscript ``R'', e.g., $\Phi_\R$ and $A_{\R\mu}$.

Hereafter, we frequently use the short-hand notation for the momentum integral,
\al{
\int_p := \int \frac{\df^d p}{\paren{2\pi}^d},
}
where $d=4-2\vep$ with an infinitesimal variable $\vep>0$.
The propagators are described as
\al{
\langle A_\mu\fn{x} A_\nu\fn{y} \rangle_0
	&=
		{\paren{ g_0^2 \mu^{2\vep} }}
		\int_p \frac{e^{ip \cdot \paren{x-y}}}{ p^2 }
		\sqbr{
			\pn{\delta_{\mu\nu}  - {p_\mu p_\nu\ov p^2} } + \xi_0{p_\mu p_\nu\ov p^2}
		}, \\
\langle \Phi\fn{x} \Phi^\dagger\fn{y} \rangle_0
	&=
		{\paren{ g_0^2 \mu^{2\vep} }}
		\int_p \frac{e^{ip \cdot \paren{x-y}}}{ p^2 + m_0^2 },
}
where the expectation value is defined in the language of path integral as
\al{
\langle {\cal O} \rangle_0
	:=
		\frac
		{  \int {\cal D}A {\cal D} \Phi {\cal D} \Phi^\dagger {\cal O} \, e^{- S_\text{free}}  }
		{  \int {\cal D}A {\cal D} \Phi {\cal D} \Phi^\dagger e^{- S_\text{free}}  },
		\label{eq:path-integral-average}
}
in which $S_\tx{free}$ denotes the free quadratic part of the action~\eqref{eq:bulk-action}.
Here, we introduce the renormalization scale $\mu$ to compensate for the physical dimension of the gauge coupling $g_0$.

The relations between the bare fields and renormalized fields are introduced,
\al{
A_\mu &= \sqrt{Z_A} A_{\R\mu},&
\Phi &= \sqrt{Z_\Phi} \Phi_{\R},&
}
where the wave function renormalization factors are parameterized as
\al{
Z_A &:= 1 + \delta Z_A,& 
Z_\Phi &:= 1 + \delta Z_\Phi.&
}
And we introduce between the bare couplings and renormalized couplings,
\al{
m_0^2 &= Z_\Phi^{-1} \paren{m^2 + \delta m^2},&
g_0 &= Z_A^{-1/2} \paren{g + \delta g},& \notag \\
\xi_0 &= 1 + \delta \xi,&
\lambda_0 &= Z_\Phi^{-2} \paren{\lambda + \delta\lambda},\label{gaugefix}&
}
where we adopt the notation for couplings where renormalized ones do not hold any subscripts, such as $m$ and $g$;
$\delta m^2$, $\delta g$, $\delta \xi$, and $\delta \lambda$ are the counterparts for the scalar mass, the gauge coupling, the gauge-fixing parameter, and the scalar self-coupling, respectively.

In the minimal-subtraction scheme, the counterparts are identified at the one-loop level as
\al{
\delta Z_A &= \frac{1}{\paren{4\pi}^2 \vep} \paren{ - \frac{g^2}{3} },&
\delta Z_\Phi &= \frac{1}{\paren{4\pi}^2 \vep} \paren{ 2g^2 },& \notag \\
\delta m^2 &= \frac{1}{\paren{4\pi}^2 \vep} \paren{ 4\lambda m^2 - g^2 m^2 },&
\delta g &= 0,& \notag \\
\delta \xi &= \frac{1}{\paren{4\pi}^2 \vep} \paren{ - \frac{g^2}{3} },&
\delta \lambda &= \frac{1}{\paren{4\pi}^2 \vep} \paren{ 10\lambda^2 + 5g^4 - {2} g^2 \lambda },
\label{eq:delta-couplings}
}
which leads to\footnote{
The one-loop relationship between $\lambda_0$ and $\lambda$ is not necessary for the following discussion.
}
\al{
m_0^2 
	&= m^2 + \frac{1}{\paren{4\pi}^2 \vep} \paren{ 4\lambda m^2 -3 g^2 m^2 }, 
	\label{eq:m_0-renomalization} \\
g_0 
	&= g + \frac{1}{\paren{4\pi}^2 \vep} \paren{\frac{g^3}{6}}, 
	\label{eq:g_0-renomalization}  \\
\lambda_0 
	&= \lambda + \frac{1}{\paren{4\pi}^2 \vep} \paren{ 10\lambda^2 + 5g^4 - {6} g^2 \lambda }.
	\label{eq:lambda_0-renomalization} 
}

\subsection{Gradient flow}

Gradient flow is an efficient method to describe a well-defined expectation value of a composite operator through diffusion into the direction of a flow time $t$, called the flow time~\cite{Narayanan:2006rf,Luscher:2009eq,Luscher:2010iy,Luscher:2011bx,Luscher:2013cpa}.\footnote{
You can refer to the materials provided by Hiroshi Suzuki for the lectures at Osaka University from November 14 to 16, 2018~\cite{Suzuki-Lecture-Osaka-1,Suzuki-Lecture-Osaka-2}.
}
For $A_\mu\fn{x}$ and $\Phi\fn{x}$,
we introduce a flow-time-dependent gauge field $B_\mu\fn{t, x}$ and a complex scalar field $\Psi\fn{t, x}$, respectively, for $t\geq0$ under the initial condition at $t=0$:
\al{
&B_\mu\fn{t=0, x} = A_\mu\fn{x},&
&\Psi\fn{t=0, x} = \Phi\fn{x},&
&\Psi^\dagger\fn{t=0, x} = \Phi^\dagger\fn{x}.
}
We introduce a flowed action:
\al{
\widetilde{S}
	&=	
		\frac{1}{g_0^2}\int_0^\infty\df t \int \df^d x 
		\br{
			\frac{1}{4} G_{\mu\nu}\fn{t,x} G_{\mu\nu}\fn{t,x}
			+\frac{\alpha_0}{2} \Pn{ \p_\mu B_\mu\fn{t,x}}^2 
			+\Pn{D_\mu \Psi\fn{t,x}}^\dagger D_\mu \Psi\fn{t,x}
			},
			\label{flowed action}
}
where $G_{\mu\nu} := \p_\mu B_\nu-\p_\nu B_\mu$ and $\alpha_0$ is a dimensionless gauge-fixing constant in the bulk $\paren{t > 0}$.
In the flowed action, we dropped the scalar mass and self-interaction terms, which include the bare parameters that contain the divergent counter terms in order to be related to those of the boundary theory at $t=0$. There is no bulk divergence that cancels such a counter term~\cite{Kadoh:2023mof,Capponi:2015ucc}.

The gradient flow equations are a kind of diffusion equation introduced as
\al{
\p_t B_\mu\fn{t,x}
	&= 
		- g_0^2  \frac{\delta \widetilde{S}}{\delta B_\mu\fn{t,x}}  \notag \\
	&= 
		\paren{ \delta_{\mu\nu} \, \p^2 -\p_\mu \p_\nu } B_\nu\fn{t,x} + \alpha_0 \, \p_\mu \p_\nu B_\nu\fn{t,x} + R^B_\mu\fn{t,x}, 
		\label{eq:GFeq-B} \\
\p_t \Psi\fn{t,x}
	&=
		- g_0^2  \frac{\delta \widetilde{S}}{\delta \Psi^\dagger\fn{t,x}} \notag \\
	&=
		\p^2 \Psi\fn{t,x} + R^\Psi\fn{t,x},
		\label{eq:GFeq-Psi} \\
\p_t \Psi^\dagger\fn{t,x}
	&=
		- g_0^2 \frac{\delta \widetilde{S}}{\delta \Psi\fn{t,x}}  \notag \\
	&=
		\p^2 \Psi^\dagger\fn{t,x} + R^{\Psi^\dagger}\fn{t,x},
		\label{eq:GFeq-Psibar}
}
with nonlinear interaction parts,
\al{
R^B_\mu
	&:=
		i \paren{ \Psi^\dagger \p_\mu \Psi - \paren{\p_\mu \Psi^\dagger} \Psi } + 2 B_\mu \Psi^\dagger \Psi, 
		\label{eq:form-R^B} \\
R^\Psi
	&:=
		-2 i \, B_\mu \p_\mu \Psi -i \paren{ \p_\mu B_\mu } \Psi - B_\mu B_\mu \Psi,
		\label{eq:form-R^Psi} \\
R^{\Psi^\dagger}
	&:=
		2 i \, B_\mu \p_\mu \Psi^\dagger +i \paren{ \p_\mu B_\mu } \Psi^\dagger - B_\mu B_\mu \Psi^\dagger,
		\label{eq:form-R^Psibar}
}
where the symbol $\delta$ represents the variation and we introduce $\p^2 := \p_\mu \p_\mu$.

We note that if all the flowed fields have fixed values at $t\to\infty$, the left-hand sides of the gradient flow equations become zero, and hence the limiting values of the flowed fields are given by the variation of the flowed action~\eqref{flowed action}. Since the flowed Higgs field has no potential in the flowed action, any constant flowed Higgs field (independent of $x$) can be a solution to the variation of the flowed action {in the limit $t\to\infty$ if the limiting value of $B_\mu$ is zero}. That is, this setup allows any value for the flowed Higgs field in the large $t$ limit {for the vanishing flowed gauge field}. It is nontrivial whether the flowed order parameter takes a nonzero value or not.

The flow equations can be formally solved as
\al{
B_\mu\fn{t,x}
	&=
		\int \df^d y 
		\br{
			\sqbr{K^{\paren{\alpha_0}}_t \paren{x-y}}_{\mu\nu} A_\nu\fn{y} + \int_0^t \df s \sqbr{K^{\paren{\alpha_0}}_{t-s} \paren{x-y}}_{\mu\nu} R^B_\nu\fn{{s}, y}
		}, 
		\label{eq:flowed-B} \\
\Psi\fn{t,x}
	&=
		\int \df^d y 
		\br{
			K_t \paren{x-y} \Phi\fn{y} + \int_0^t \df s K_{t-s} \paren{x-y} R^\Psi\fn{{s}, y}
		},
		\label{eq:flowed-Psi}  \\
\Psi^\dagger\fn{t,x}
	&=
		\int \df^d y 
		\br{
			K_t \paren{x-y} \Phi^\dagger\fn{y} + \int_0^t \df s K_{t-s} \paren{x-y} R^{\Psi^\dagger}\fn{{s}, y}
		},
		\label{eq:flowed-Psibar} 
}
where the following heat kernels are incorporated:
\al{
\sqbr{K^{\paren{\alpha_0}}_t \paren{x}}_{\mu\nu}
	&:=
		\int_p  e^{ip \cdot x }  
		\sqbr{
			\paren{ \delta_{\mu\nu}  - {p_\mu p_\nu\ov p^2} } e^{-t p^2} + {p_\mu p_\nu\ov p^2} e^{- \alpha_0 t p^2}
		}, 
		\label{eq:Kt-form-gauge-general} \\
K_t \paren{x}
	&:=
		\int_p { e^{ip \cdot x } }
		 e^{-t p^2}.
		\label{eq:Kt-form-scalar}
}
Concrete forms of the classical formal solutions in Eqs.~\eqref{eq:flowed-B}, \eqref{eq:flowed-Psi}, and \eqref{eq:flowed-Psibar}
can be derived in a recursive way,
depending on how many times each of the corresponding nonlinear interaction terms in Eqs.~\eqref{eq:form-R^B}, \eqref{eq:form-R^Psi}, and \eqref{eq:form-R^Psibar} are incorporated and also on the order of the incorporation.
We adopt the following short-hand notation to reduce the complexity of the following description:
\al{
\wt{B}_\mu\fn{t,x}
	&:=
		\int \df^d y 
			\sqbr{K_t \paren{x-y}}_{\mu\nu} A_\nu\fn{y}, \notag \\
\wt{\Psi}\fn{t,x}
	&:=
		\int \df^d y 
			K_t \paren{x-y} \Phi\fn{y}, \label{eq:fields_LO} \\
\wt{\Psi}^\dagger\fn{t,x}
	&:=
		\int \df^d y 
			K_t \paren{x-y} \Phi^\dagger\fn{y},\notag	
}
where
\begin{align}
\sqbr{K_t \paren{x}}_{\mu\nu} &:=
\sqbr{K^{\paren{\alpha_0=1}}_t \paren{x}}_{\mu\nu}
		=
		\int_p { e^{ip \cdot x } } \delta_{\mu\nu}
		 e^{-t p^2}.
\end{align}
The quantum two-point correlation functions of the flowed gauge bosons and the flowed Higgs bosons are estimated through the operation in path integral in Eq.~\eqref{eq:path-integral-average} as
\al{
\langle B_\mu\fn{t,x} B_\nu\fn{s,y} {\cal X} \rangle_0, \qquad
\langle \Psi\fn{t,x} \Psi^\dagger\fn{s,y} {\cal X} \rangle_0,
}
where ${\cal X}$ symbolically represents possible contributions from perturbative expansions of $e^{- S_\text{int}}$, where
$S_\text{int}$ is the interaction part of the original action in Eq.~\eqref{eq:bulk-action}:
${\cal X}$ is unity for leading-order~(LO) calculations, while
${\cal X}$ can become nontrivial (non-unity) at a loop level.
It is straightforward to derive the LO result of the two-point functions:
\al{
\left< B_\mu\fn{t,x} B_\nu\fn{s,y} \right>_\text{LO}
	&=
		\left< \wt{B}_\mu\fn{t,x} \wt{B}_\nu\fn{s,y} \right>_0 \notag \\
	&=
		\left<
		\int \df^d z\sqbr{K_t^{(\alpha_0)}\fn{x-z}}_{\mu\rho} A_\rho\fn{z}
		\int \df^d \wt{z}\sqbr{K_s^{(\alpha_0)}\fn{y-\wt{z}}}_{\nu\sigma} A_\sigma\fn{\wt{z}}
		\right>_0 \notag \\
	&=
		{\paren{ g_0^2 \mu^{2\vep} }}
		\int_p \frac{ e^{ip \cdot \paren{x-y} } }{ p^2 }
		\sqbr{
			\paren{ \delta_{\mu\nu}  - {p_\mu p_\nu\ov p^2} } e^{- \paren{t+s} p^2} + \xi_0 {p_\mu p_\nu\ov p^2} e^{- \alpha_0 \paren{t+s} p^2}
		},\label{eq:Boson-2PTform-LO} \\
\left< \Psi\fn{t,x} \Psi^\dagger\fn{s,y} \right>_\text{LO}
	&=
		\left< \wt{\Psi}\fn{t,x} \wt{\Psi}^\dagger\fn{s,y} \right>_0 \notag \\
	&=
		\left<
		\int \df^d z K_t\fn{x-z} \Phi\fn{z}
		\int \df^d \wt{z} K_s\fn{y-\wt{z}} \Phi^\dagger\fn{\wt{z}}
		\right>_0 \notag \\
	&=
		{\paren{ g_0^2 \mu^{2\vep} }}
		\int_p \frac{ e^{ip \cdot \paren{x-y} - \paren{t+s} p^2 } }{ p^2 + m_0^2 },
		\label{eq:Psi-2PTform-LO}
}
where we have used the definition of the operation~\eqref{eq:path-integral-average},
Wick's theorem, and the form of the heat kernels in Eqs.~\eqref{eq:Kt-form-gauge-general} and \eqref{eq:Kt-form-scalar}.\footnote{
{
Currently, $\left< \Psi\fn{t,x} \Psi^\dagger\fn{s,y} \right> = \left< \Psi^\dagger\fn{s,y} \Psi\fn{t,x} \right>$ is realized since we consider an Abelian gauge theory. For a flowed complex scalar field obeying the fundamental representation of the $SU(N)$ non-Abelian gauge theory $\Psi_a$\,($a=1,2,...,N$), the corresponding order parameter should be
$\left< \sum_{a=1}^N \Psi_a\fn{t,x} \Psi_a^\dagger\fn{s,y} \right> = \left< \sum_{a=1}^N \Psi_a^\dagger\fn{s,y} \Psi_a\fn{t,x} \right>$.
}
}

As we will soon see, the two-point function of the flowed massless gauge boson $B_\mu$ becomes finite without a wave function renormalization of the flowed field, while that of the flowed complex scalar field $\Psi$ still contains a divergent part.
To eliminate this part systematically, we introduce a wave function renormalization factor for $\Psi$ as follows:
\al{
\Psi &= \sqrt{{\cal Z}_\Psi} \Psi_\R,
\label{eq:bulk-scalar-counterpart-formal}
}
with the counterpart $\delta {\cal Z}_\Psi$ defined as
\al{
{\cal Z}_\Psi &:= 1 + \delta {\cal Z}_\Psi.
\label{eq:bulk-scalar-counterpart}
}

Finally, in addition to Eq.~\eqref{eq:fields_LO}, we introduce the following symbols for later convenience:
\al{
{{\cal B}}_\mu\fn{u,z}
	&:=
		\int \df^d w \int_0^u \df u' K_{u-u'}\fn{z-w}
		\sqbr{  i \paren{ \wt{\Psi}^\dagger \p_\mu \wt{\Psi} - \paren{\p_\mu \wt{\Psi}^\dagger} \wt{\Psi} }  }_{\fn{u', w}}, \notag \\
{{\cal P}}\fn{u,z}
	&:=
		\int \df^d w \int_0^u \df u' K_{u-u'}\fn{z-w}
		\sqbr{
		-2 i \, \wt{B}_\mu \p_\mu \wt{\Psi} -i \paren{ \p_\mu \wt{B}_\mu } \wt{\Psi} 
		}_{\fn{u', w}}, 		\label{eq:NLO-shorthand}\\
{{\cal P}}^\dagger\fn{u,z}
	&:=
		\int \df^d w \int_0^u \df u' K_{u-u'}\fn{z-w}
		\sqbr{
		2 i \, \wt{B}_\mu \p_\mu \wt{\Psi}^\dagger +i \paren{ \p_\mu \wt{B}_\mu } \wt{\Psi}^\dagger 
		}_{\fn{u', w}},\notag
}
where we introduced the short-hand notation for a function $f$ of the flowed fields,
\al{
	f\fn{B_\mu\fn{u,z},\Psi\fn{u,z},\Psi^\dagger\fn{u,z}}
	=
	\sqbr{f\fn{B_\mu,\Psi,\Psi^\dagger}}_{\fn{u,z}}.
	\label{eq:non-linear-notation}
}

\section{UV finiteness of the gauge two-point function}\label{3}

\begin{figure}[t]
\centering
\includegraphics[width=0.8\textwidth]{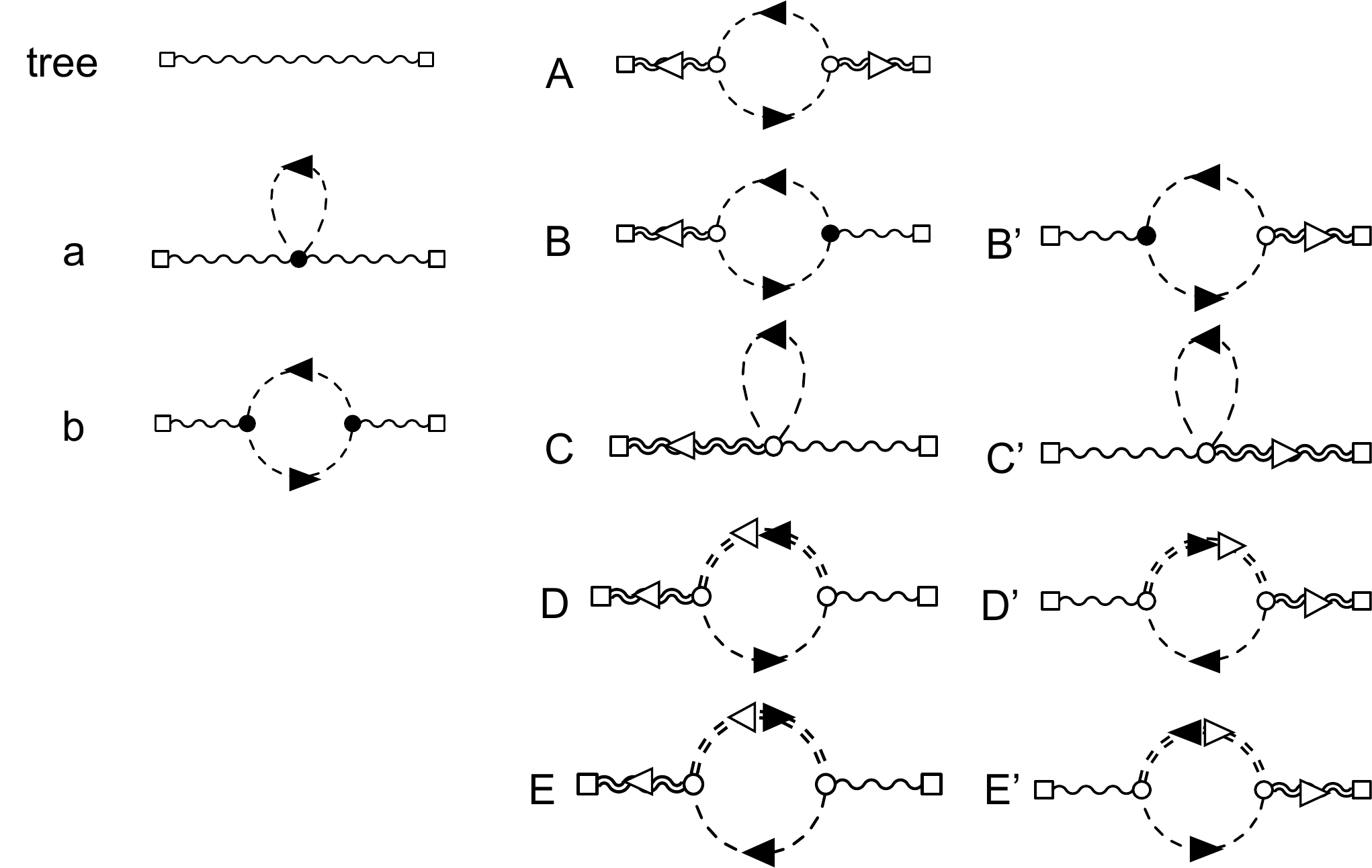}
\caption{
All the configurations at the one-loop level for the gauge-boson two-point function are shown in the language of the flow Feynman diagram~\cite{Luscher:2011bx}.
In each flow Feynman diagram, the left and right squares indicate the spacetime positions of $\paren{t,x}$ and $\paren{s,y}$, respectively.
The double solid and wavy lines are the flow propagators of the scalar and vector fields, respectively, i.e., \ their flow-time evolutions with the nonlinear terms~\eqref{eq:GFeq-B}--\eqref{eq:GFeq-Psibar}, where both of the endpoints are located in the bulk (at a flow time $t>0$).
Black arrows and white-out arrows indicate the directions of the Higgs particle number and of the flow time, respectively.
Each white circle, which is called a flow vertex, indicates a branching in the bulk due to one of the possible nonlinear interactions in the flow equations.
Each black dot shows an ordinary vertex at the boundary ($t=0$).
All the diagrams are drawn by the package~{\tt feynMF}~\cite{Ohl:1995kr}.
}
\label{Fig:diagram-gaugeboson2pt}
\end{figure}

As we saw in Eqs.~\eqref{eq:Boson-2PTform-LO} and \eqref{eq:Psi-2PTform-LO}, at the LO, the two-point functions of the gauge field and the Higgs field are
constructed by the leading-order solutions, shown in Eq.~\eqref{eq:fields_LO}, of the gradient flow equations~\eqref{eq:GFeq-B}--\eqref{eq:GFeq-Psibar}.
The next-to-leading two-point functions of the flowed fields can be calculated in a similar way, namely,
by taking an expectation value of two sequential approximation solutions [refer to Eq.~\eqref{eq:path-integral-average}, also Eqs.~\eqref{eq:flowed-B}--\eqref{eq:flowed-Psibar}]. Here, possible configurations are classified into two categories at the one-loop level:
\begin{itemize}
\item[(i)] It involves at least one field that is evolved nonlinearly in its flow equation.
\item[(ii)] It does not involve fields that are evolved nonlinearly in their flow equations, while a one-loop structure is observed at the boundary with the zero value of the flow time.
\end{itemize}
{
The gauge two-point function does not have a tadpole diagram in the symmetric phase. For the scalar two-point function, we will explicitly show that the tadpole diagrams become zero.}
At the one-loop level, in the language of the flow Feynman diagram~\cite{Luscher:2011bx}, possible configurations contributing to the gauge-boson two-point function are summarized as 11 diagrams as shown in Fig.~\ref{Fig:diagram-gaugeboson2pt}. In the following loop calculation, we will set the two gauge-fixing parameters as
\al{
\alpha_0 = \xi_0 = 1,
\label{eq:NLO-GF}
}
which is a Feynman gauge-like configuration in both the bulk (at a flow time $t>0$) and the boundary ($t=0$); see also \eref{gaugefix} and \eref{eq:delta-couplings} in the tree calculation.

\subsection{Configurations}

Adopting the short-hand notations in Eqs.~\eqref{eq:fields_LO} and \eqref{eq:NLO-shorthand} allows us to describe the details of each configuration efficiently.
First, we write down the five configurations classified into Category (i), which involve at least one field that is evolved nonlinearly in its flow equation:
\al{
\left< B_\mu\fn{t,x} B_\nu\fn{s,y} \right>_\text{A}
	:=&\,
		\left< {\cal B}_\mu\fn{t,x} {\cal B}_\nu\fn{s,y} \right>_{\text{connected}} \notag \\
	=& \,
		\paren{g_0^4 \mu^{4\vep}}
		\int_{p,\ell} \int_0^t \df u \int_0^s \df u'
		\frac
		{\paren{p+2\ell}_\mu \paren{p+2\ell}_\nu e^{ip\cdot\paren{x-y} - \paren{s+t}p^2}}
		{\paren{\ell^2+m_0^2}\sqbr{\paren{p+\ell}^2+m_0^2}}
		e^{-2\paren{u+u'}\paren{\ell^2 {+} p\cdot \ell}},
		\\
\left< B_\mu\fn{t,x} B_\nu\fn{s,y} \right>_\text{B}
	:=&\,
		\left< 
			{\cal B}_\mu\fn{t,x}
			\br{
		-\frac{1}{g_0^2} \int \df^d X \sqbr{ i A_\sigma \paren{ \Phi^\dagger \p_\alpha \Phi 
				- \paren{\p_\alpha \Phi^\dagger} \Phi } }_{\fn{X}
			}}
			\wt{B}_\nu\fn{s,y}
		\right>_{\text{connected}} \notag \\
	=& \,
		\paren{-g_0^4 \mu^{4\vep}}
		\int_{p,\ell} \int_0^t \df u
		\frac
		{\paren{p+2\ell}_\mu \paren{p+2\ell}_\nu e^{ip\cdot\paren{x-y} - \paren{s+t}p^2}}
		{p^2 \paren{\ell^2+m_0^2}\sqbr{\paren{p+\ell}^2+m_0^2}}
		e^{-2u\paren{\ell^2 {+} p\cdot \ell}},
		\\
\left< B_\mu\fn{t,x} B_\nu\fn{s,y} \right>_\text{C}
	:=&\,
		\left< 
			\int \df^d z \int_0^t \df u K_{t-u} \fn{x-z} \sqbr{ 2 \wt{B}_\mu  \wt{\Psi}^\dagger \wt{\Psi} }_{\fn{u,z}}
			\wt{B}_\nu\fn{s,y}
		\right>_{\text{connected}} \notag \\
	=& \,
		\paren{2 g_0^4 \mu^{4\vep}} \int_{p,\ell} \int_0^t \df u
		\frac{\delta_{\mu\nu} \, e^{ip\cdot\paren{x-y} - \paren{s+t}p^2}}
		{p^2 \paren{\ell^2 + m_0^2}}
		e^{-2 u \ell^2},
		\\
\left< B_\mu\fn{t,x} B_\nu\fn{s,y} \right>_\text{D}
	:=&\,
		\left< 
			\int \df^d z \int_0^t \df u K_{t-u} \fn{x-z}
			\br{ i \sqbr{ \wt{\Psi}^\dagger \p_\mu {\cal P} - \paren{\p_\mu \wt{\Psi}^\dagger} {\cal P} }_{\fn{u,z}} }
			\wt{B}_\nu\fn{s,y}
		\right>_{\text{connected}} \notag \\
	=& \,
		\paren{-g_0^4 \mu^{4\vep}}
		\int_{p,\ell} \int_0^t \df u \int_0^u \df\widehat{u}
		\frac
		{\paren{2\ell-p}_\mu \paren{2\ell-p}_\nu e^{ip\cdot\paren{x-y} - \paren{s+t}p^2}}
		{p^2 \sqbr{\paren{p-\ell}^2+m_0^2}}
		e^{-2u\paren{\ell^2 - p\cdot \ell} -2 \widehat{u}\paren{p^2 - p\cdot \ell}},
		\\
\left< B_\mu\fn{t,x} B_\nu\fn{s,y} \right>_\text{E}
	:=&\,
		\left< 
			\int \df^d z \int_0^t \df u K_{t-u} \fn{x-z}
			\br{ i \sqbr{ {\cal P}^\dagger \p_\mu \wt{\Psi} - \paren{\p_\mu {\cal P}^\dagger} \wt{\Psi}  }_{\fn{u,z}} }
			\wt{B}_\nu\fn{s,y}
		\right>_{\text{connected}} \notag \\
	=& \,
		\paren{-g_0^4 \mu^{4\vep}}
		\int_{p,\ell} \int_0^t \df u \int_0^u \df\widehat{u}
		\frac
		{\paren{2\ell-p}_\mu \paren{2\ell-p}_\nu e^{ip\cdot\paren{x-y} - \paren{s+t}p^2}}
		{p^2 \sqbr{\paren{p-\ell}^2+m_0^2}}
		e^{-2u\paren{\ell^2 - p\cdot \ell} -2 \widehat{u}\paren{p^2 - p\cdot \ell}},
}
where the labels (A) to (E) tell us how they relate to the flowed Feynman diagrams in Fig.~\ref{Fig:diagram-gaugeboson2pt},
{and the label ``connected" indicates the operation required to drop off disconnected configurations from Eq.~\eqref{eq:path-integral-average}.}
Here, $p$ and $\ell$ represent the physical momentum flowing from the point $y$ to $x$ and a loop momentum, respectively.
To describe the configuration (B) simply, we introduced a similar notation for the fields on the boundary as introduced in Eq.~\eqref{eq:non-linear-notation}.
Note that the result in (D) is the same as that in (E).
The remaining four contributions classified into Category (i), namely (B$'$) to (E$'$), are simply obtained by the parameter replacement as
\al{
\left< B_\mu\fn{t,x} B_\nu\fn{s,y} \right>_\text{B$'$}
	&=
		{\left(\left< B_\nu\fn{s,y} B_\mu\fn{t,x} \right>_\text{B}\right)^*},&
\left< B_\mu\fn{t,x} B_\nu\fn{s,y}  \right>_\text{C$'$}
	&=
		{\left(\left< B_\nu\fn{s,y} B_\mu\fn{t,x} \right>_\text{C}\right)^*}, \notag \\
\left< B_\mu\fn{t,x} B_\nu\fn{s,y} \right>_\text{D$'$}
	&=
		{\left(\left< B_\nu\fn{s,y} B_\mu\fn{t,x} \right>_\text{D}\right)^*},&
\left< B_\mu\fn{t,x} B_\nu\fn{s,y} \right>_\text{E$'$}
	&=
		{\left(\left< B_\nu\fn{s,y} B_\mu\fn{t,x} \right>_\text{E}\right)^*}.
}

Two configurations are classified into Category (ii) ``with a loop on the boundary,'' where their corresponding flow Feynman diagrams labeled as (a) and (b) are shown in Fig.~\ref{Fig:diagram-gaugeboson2pt}:
\al{
\left< B_\mu\fn{t,x} B_\nu\fn{s,y} \right>_\text{a}
	:=&\,
		\left< 
			\wt{B}_\mu\fn{t,x}
		\br{
			-\frac{1}{g_0^2} \int \df^d X \sqbr{ A_\alpha A_\alpha \Phi^\dagger \Phi }_{\fn{X}}
			}
			\wt{B}_\nu\fn{s,y}
		\right>_{\text{connected}} \notag \\
	=& \,
		\paren{-2g_0^4 \mu^{4\vep}}
		\int_p
		\frac{\delta_{\mu\nu} \, e^{ip\cdot\paren{x-y} - \paren{s+t}p^2}}
		{\paren{p^2}^2}
		\int_\ell \frac{1}{\ell^2 + m_0^2},
		\\
\left< B_\mu\fn{t,x} B_\nu\fn{s,y} \right>_\text{b}
	:=& \,		
		\Bigg<
		\wt{B}_\mu\fn{t,x}
		\frac{1}{2!}
		\br{
			-
			\frac{1}{g_0^2}
			\int \df^d X \sqbr{  i A_\alpha \paren{ \Phi^\dagger \p_\alpha \Phi - \paren{\p_\alpha \Phi^\dagger} \Phi  }  }_{\fn{X}}
		} \notag \\
	&\quad \times
		\br{
			-
			\frac{1}{g_0^2}
			\int \df^d Y \sqbr{  i A_\beta \paren{ \Phi^\dagger \p_\beta \Phi - \paren{\p_\beta \Phi^\dagger} \Phi  } }_{\fn{Y}} 
		}
		\wt{B}_\mu\fn{s,y}
		\Bigg>_{\text{connected}} \notag \\
	=&\,
		\paren{g_0^4 \mu^{4\vep}}
		\int_{p,\ell}
		\frac{\delta_{\mu\nu} \, e^{{-} ip\cdot\paren{x-y} - \paren{s+t}p^2}}
		{\paren{p^2}^2}
		\frac{\paren{2\ell-p}_\mu \paren{2\ell-{p}}_\nu}
		{\paren{\ell^2 + m_0^2} \sqbr{ \paren{\ell-p}^2 + m_0^2 }}.
}
From each of the above results, we recognize that the singularity for the limit $y \to x$ is cured due to the smearing through the gradient flow for $t, s>0$.

We can identify the UV poles due to the integrals of the loop momentum $\ell$ before the inverse Fourier transforms.
Here, (at least) couplings should be renormalized even in the gradient flow.
Now, our discussion is at the one-loop level, and thus we should use the one-loop relationships summarized in Eqs.~\eqref{eq:m_0-renomalization}, \eqref{eq:g_0-renomalization}, and \eqref{eq:lambda_0-renomalization} for the LO part (such as in Eq.~\eqref{eq:BB-LO-div} below),
while the following LO relationships are sufficient for the one-loop part:
\al{
m_0^2 &\LOeq m^2,&
g_0 &\LOeq g,&
\lambda_0 &\LOeq \lambda.&
}
Each expression is written in the renormalized couplings, where we ignore higher-order parts in multiplicative computations:
\al{
\left.\left< B_\mu\fn{t,x} B_\nu\fn{s,y} \right>_\text{A}\right|_\text{UV-div}
	&=
		0, 
		\label{eq:BB-A-div} \\
\left.\left< B_\mu\fn{t,x} B_\nu\fn{s,y} \right>_\text{B}\right|_\text{UV-div}
	&=
		\left.\left< B_\mu\fn{t,x} B_\nu\fn{s,y} \right>_\text{B$'$}\right|_\text{UV-div}
	=
		\int_p \frac{\delta_{\mu\nu}\,e^{ip\cdot\paren{x-y} -\paren{t+s}p^2}}{ p^2 }
		\paren{  - \frac{ g^4 }{2\paren{4\pi}^2} \frac{1}{\vep}   }, \\
\left.\left< B_\mu\fn{t,x} B_\nu\fn{s,y} \right>_\text{C}\right|_\text{UV-div}
	&=
		\left.\left< B_\mu\fn{t,x} B_\nu\fn{s,y} \right>_\text{C$'$}\right|_\text{UV-div}
	=
		\int_p \frac{\delta_{\mu\nu}\,e^{ip\cdot\paren{x-y} -\paren{t+s}p^2}}{ p^2 }
		\paren{   \frac{ g^4 }{\paren{4\pi}^2} \frac{1}{\vep}   }, \\
\left.\left< B_\mu\fn{t,x} B_\nu\fn{s,y} \right>_\text{D}\right|_\text{UV-div}
	&=
		\left.\left< B_\mu\fn{t,x} B_\nu\fn{s,y} \right>_\text{D$'$}\right|_\text{UV-div} \notag  \\
= \left.\left< B_\mu\fn{t,x} B_\nu\fn{s,y} \right>_\text{E}\right|_\text{UV-div}
	&=
		\left.\left< B_\mu\fn{t,x} B_\nu\fn{s,y} \right>_\text{E$'$}\right|_\text{UV-div}
	=
		\int_p \frac{\delta_{\mu\nu}\,e^{ip\cdot\paren{x-y} -\paren{t+s}p^2}}{ p^2 }
		\paren{  - \frac{ g^4 }{4\paren{4\pi}^2} \frac{1}{\vep}   }, \\
\left.\left< B_\mu\fn{t,x} B_\nu\fn{s,y} \right>_\text{a$+$b}\right|_\text{UV-div}
	&=
		\int_p \frac{e^{ip\cdot\paren{x-y} -\paren{t+s}p^2}}{ p^2 }
		\sqbr{  \frac{ g^4 }{\paren{4\pi}^2} \frac{1}{\vep}  \paren{-\frac{1}{3}\delta_{\mu\nu} + \frac{1}{3} \frac{p_\mu p_\nu}{p^2} }  }, \\
\left.\left< B_\mu\fn{t,x} B_\nu\fn{s,y} \right>_\text{LO}\right|_\text{UV-div}
	&=
		\left.
		\paren{ g_0^2 \mu^{2\vep} }
		\int_p \frac{ e^{ip \cdot \paren{x-y} -\paren{s+t}p^2} }{ p^2 }
		\sqbr{
			\paren{ \delta_{\mu\nu} - \frac{p_\mu p_\nu}{p^2} } 
			 + \xi_0 \frac{p_\mu p_\nu}{p^2} 
		}
		\right|_\text{UV-div} \notag \\
	&=
		\int_p \frac{ e^{ip \cdot \paren{x-y} -\paren{s+t}p^2} }{ p^2 }
		\br{
			\frac{ g^4 }{3\paren{4\pi}^2} \frac{1}{\vep} 
			\sqbr{
				\paren{ \delta_{\mu\nu} - \frac{p_\mu p_\nu}{p^2} } 
				 +  \frac{p_\mu p_\nu}{p^2}
			}
			-
			\frac{ g^4 }{3\paren{4\pi}^2} \frac{1}{\vep} 
			\frac{p_\mu p_\nu}{p^2}
		},
		\label{eq:BB-LO-div}
}
where we used the one-loop relationships between the bare couplings and renormalized ones, especially $\xi_0=1+\delta\xi$ with $\delta\xi$ given in Eq.~\eqref{eq:delta-couplings}.
Note that the first and second terms of Eq.~\eqref{eq:BB-LO-div} originate from $g_0^2$ and $\xi_0$, respectively.
It is straightforward to sum over the contributions~\eqref{eq:BB-A-div}--\eqref{eq:BB-LO-div} and to reach the conclusion,
\al{
\left.\left< B_\mu\fn{t,x} B_\nu\fn{s,y} \right>_\text{total}\right|_\text{UV-div} = 0,
}
which means that no {extra} wave function renormalization is necessary to regularize the gauge-boson two-point function,
as expected via the ($d+1$)-dimensional Becchi-Rouet-Stora-Tyutin (BRST) symmetry that ensures the cancellation~\cite{Luscher:2010iy,Luscher:2011bx}.

\section{Higgs two-point function as flowed order parameter}\label{4}

\begin{figure}[t]
\centering
\includegraphics[width=0.7\textwidth]{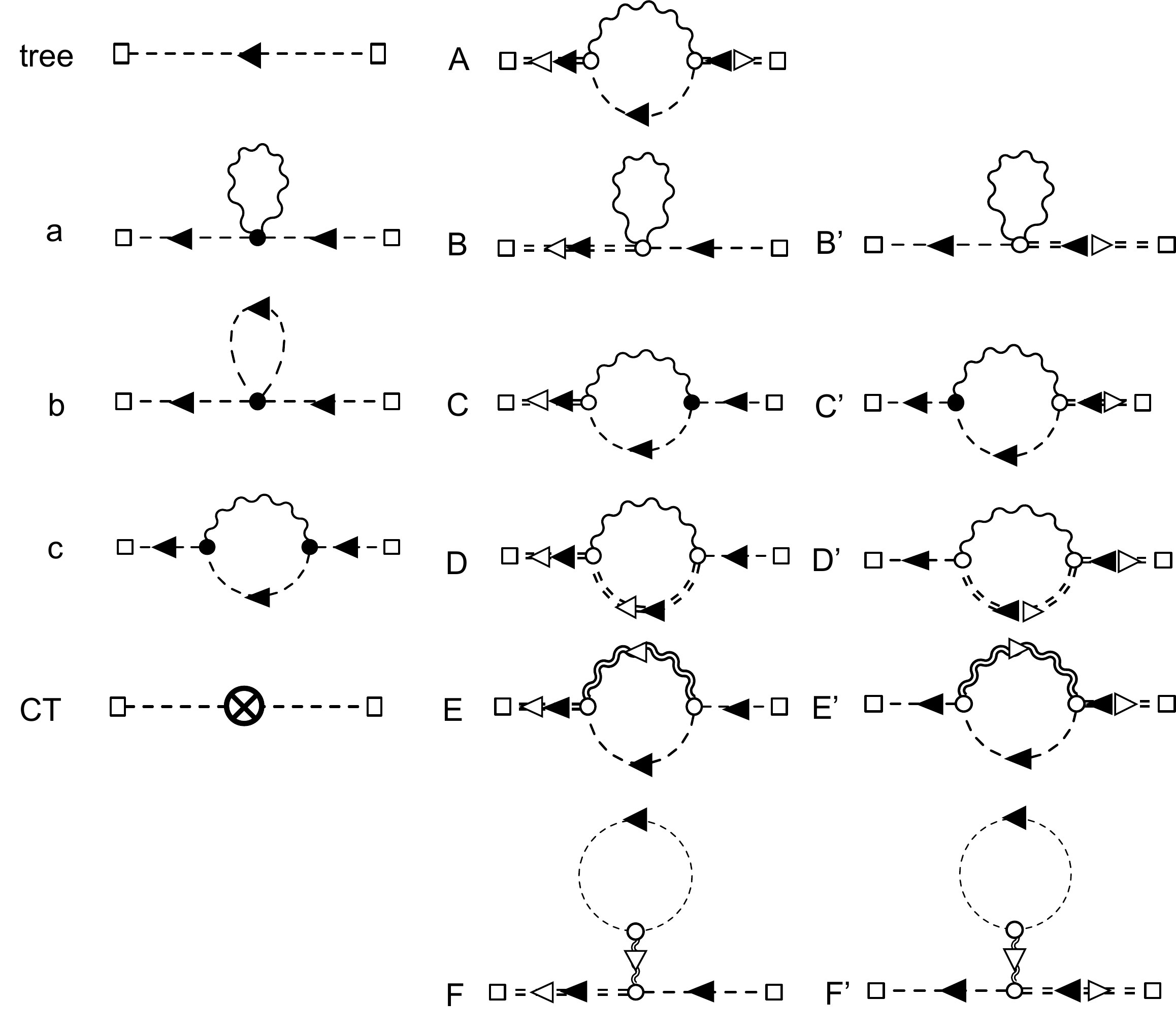}
\caption{
All the configurations at the one-loop level for the Higgs two-point function are shown in the language of the flow Feynman diagram~\cite{Luscher:2011bx}.
In each flow Feynman diagram, the left and right squares indicate the spacetime positions of $\paren{t,x}$ and $\paren{s,y}$, respectively.
The double solid and wavy lines are flow propagators of the scalar and vector fields, respectively, i.e., \ their flow time evolutions with the nonlinear terms~\eqref{eq:GFeq-B}--\eqref{eq:GFeq-Psibar}, where both endpoints are located in the bulk (at a flow time of ($t>0$)).
Black arrows and white-out arrows indicate the directions of the Higgs particle number and of the flow time, respectively.
Each white circle, which is the flow vertex, indicates a branching in the bulk due to one of the possible nonlinear interactions in the flow equations.
Each black dot shows an ordinary vertex on the boundary ($t=0$).
All the diagrams are drawn by the package~{\tt feynMF}~\cite{Ohl:1995kr}.
}
\label{Fig:diagram-Higgs2pt}
\end{figure}

Next, we focus on the gauge-invariant two-point function of the flowed Higgs field at the one-loop level.
Again, the primary calculation method is the same as in the previous section, and the same gauge fixings in \eref{gaugefix} and $\alpha_0=1$ are used. Also, we take into account the bulk wave function renormalization~\eqref{eq:bulk-scalar-counterpart-formal},
where for the current one-loop calculation, the following relation is used (except for the counter term; see Eqs.~\eref{eq:Psi-2PTform-CT}  and \eref{wave_renormalize} below):
\al{
\Psi\fn{t,x} = \Psi_\R\fn{t,x}.
}


\subsection{Configurations}

The following five configurations involve at least one field that is evolved nonlinearly in its flow equation [in Category (i)]:
\al{
\left< \Psi\fn{t,x} \Psi^\dagger\fn{s,y} \right>_\text{A}
	:=&\,
		\left<
			{\cal P}\fn{t,x} {\cal P}^\dagger\fn{s,y}
		\right>_{\text{connected}}
		\notag \\
	=&\,
		\paren{g_0^4 \mu^{4\vep}}
		\int_0^t \df u \int_0^s \df\wt{u} \int_{p,\ell} \frac{\paren{p+\ell}^2}{ \paren{p-\ell}^2 \paren{\ell^2 + m_0^2} }
		e^{+i p\cdot\paren{x-y} -\paren{t+s}p^2 -2\paren{u+\wt{u}}\ell^2 +2\paren{u+\wt{u}}\ell\cdot p},
	\label{eq:Psi-2PTform-A} \\
\left< \Psi\fn{t,x} \Psi^\dagger\fn{s,y} \right>_\text{B}
	:=&\,
		\left<
		\int \df^d z \int_0^t \df u K_{t-u} \fn{x-z} \sqbr{ - \wt{B}_\mu \wt{B}_\mu \wt{\Psi} }_{\fn{u,z}}
		\wt{\Psi}\fn{s,y}
		\right>_{\text{connected}} \notag \\
	=&\,
		\paren{- g_0^4 \mu^{4\vep}}
		\int_p \frac{e^{+i p\cdot\paren{x-y} -\paren{t+s}p^2}}{p^2 + m_0^2}
		\int_0^t \df u \int_\ell \frac{d}{\ell^2} e^{-2u \ell^2},
	\label{eq:Psi-2PTform-B} \\
\left< \Psi\fn{t,x} \Psi^\dagger\fn{s,y} \right>_\text{C}
	:=&\,
		\Bigg<
		{\cal P}\fn{t,x}
		\wt{\Psi}^\dagger\fn{s,y}
		\br{
			-\frac{1}{g_0^2} \int \df^d X \sqbr{ i A_\alpha \paren{ \Phi^\dagger \p_\alpha \Phi - \paren{\p_\alpha \Phi^\dagger} \Phi } }_{\fn{X}}
		}
		\Bigg>_{\text{connected}}, \notag \\
	=&\,
		\paren{g_0^4 \mu^{4\vep}}
		\int_p \frac{e^{+i p\cdot\paren{x-y} -\paren{t+s}p^2}}{p^2 + m_0^2}
		\int_0^t \df u \int_\ell \frac{ \paren{p+\ell}^2 }{ \paren{p-\ell}^2 \paren{\ell^2 + m_0^2} } e^{-2u \paren{\ell^2 - \ell\cdot p}},
	\label{eq:Psi-2PTform-C} \\
\left< \Psi\fn{t,x} \Psi^\dagger\fn{s,y} \right>_\text{D}
	:=&\,
		\Bigg<
		\int \df^d z \int_0^t \df u K_{t-u} \fn{x-z} \sqbr{ -2 i \, \wt{{B}}_\mu \p_\mu {\cal P} -i \paren{ \p_\mu \wt{{B}}_\mu } {\cal P} }_{\fn{u,z}}
		\wt{\Psi}^\dagger\fn{s,y}
		\Bigg>_{\text{connected}} \notag \\
	=&\,
		\paren{g_0^4 \mu^{4\vep}}
		\int_p \frac{e^{+i p\cdot\paren{x-y} -\paren{t+s}p^2}}{p^2 + m_0^2}
		\int_0^t \df u \int_0^u \df\wt{u}
		\int_\ell \frac{ \paren{\ell+p}^2 }{ \paren{\ell-p}^2 } e^{ -2u\paren{\ell^2 - \ell\cdot p} -2\wt{u}\paren{p^2 - \ell\cdot p}},
	\label{eq:Psi-2PTform-D} \\
\left< \Psi\fn{t,x} \Psi^\dagger\fn{s,y} \right>_\text{E}
	:=&\,
		\Bigg<
		\int \df^d z \int_0^t \df u K_{t-u} \fn{x-z} \sqbr{ -2 i \, {{\cal B}}_\mu \p_\mu \wt{\Psi} -i \paren{ \p_\mu {{\cal B}}_\mu } \wt{\Psi} }_{\fn{u,z}}
		\wt{\Psi}^\dagger\fn{s,y}
		\Bigg>_{\text{1PI}} \notag \\
	=&\,
		\paren{-g_0^4 \mu^{4\vep}}
		\int_p \frac{e^{+i p\cdot\paren{x-y} -\paren{t+s}p^2}}{p^2 + m_0^2}
		\int_0^t \df u \int_0^u \df\wt{u} \int_\ell \frac{ \paren{\ell+p}^2 }{ \ell^2 + m_0^2 }
		e^{ -2u \ell^2 + 2 \paren{u-\wt{u}} \ell\cdot p }, 
	\label{eq:Psi-2PTform-E} \\
{\left< \Psi\fn{t,x} \Psi^\dagger\fn{s,y} \right>_\text{F}}
	:=&\,
		\Bigg<
		\int \df^d z \int_0^t \df u K_{t-u} \fn{x-z} \sqbr{ -2 i \, {{\cal B}}_\mu \p_\mu \wt{\Psi} -i \paren{ \p_\mu {{\cal B}}_\mu } \wt{\Psi} }_{\fn{u,z}}
		\wt{\Psi}^\dagger\fn{s,y}
		\Bigg>_{\text{tadpole}} \notag \\
	=&\,
		\paren{-g_0^4 \mu^{4\vep}}
		\int_p \frac{e^{+i p\cdot\paren{x-y} -\paren{t+s}p^2}}{p^2 + m_0^2}
		\int_0^t du \int_0^u d\wt{u} \int_\ell \frac{ 4 p_\mu \ell_\mu }{ \ell^2 + m_0^2 }
		e^{ -2\wt{u} \ell^2  },
	\label{eq:Psi-2PTform-F}
}
where the corresponding flow Feynman diagrams are shown in Fig.~\ref{Fig:diagram-Higgs2pt}.
{In Eqs.~\eqref{eq:Psi-2PTform-E} and \eqref{eq:Psi-2PTform-F},
we introduced the labels ``1PI" and ``tadpole" for discriminating the one-particle irreducible~(1PI) part and the tadpole part.
Note that they both are categories in the connected configurations.}

The three configurations, (a) to (c) in Fig.~\ref{Fig:diagram-Higgs2pt}, do not involve fields evolved nonlinearly in their flow equations,
while one-loop structures are observed at the boundary with the zero value of the flow time [in Category (ii)]:
\al{
\left< \Psi\fn{t,x} \Psi^\dagger\fn{s,y} \right>_\text{a}
	&:=
		\Bigg<
		\wt{\Psi}\fn{t,x} \wt{\Psi}^\dagger\fn{s,y}
		\br{
			-\frac{1}{g_0^2}
			\int \df^d X \sqbr{  A_\alpha A_\alpha \Psi^\dagger \Psi }_{\fn{X}}
		}
		\Bigg>_{\text{connected}} \notag \\
	&\,\,=
		\paren{-g_0^4 \mu^{4\vep}}
		\int_p \frac{e^{+i p\cdot\paren{x-y} -\paren{t+s}p^2}}{ \paren{p^2 + m_0^2}^2 }
		\int_\ell \frac{d}{ \ell^2 + \mu_A^2 },
	\label{eq:Psi-2PTform-a} \\
\left< \Psi\fn{t,x} \Psi^\dagger\fn{s,y} \right>_\text{b}
	&:=
		\Bigg<
		\wt{\Psi}\fn{t,x} \wt{\Psi}^\dagger\fn{s,y}
		\br{
			-\frac{\lambda_0}{g_0^4}
			\int \df^d X \sqbr{  \Psi^\dagger \Psi \Psi^\dagger \Psi  }_{\fn{X}}
		}
		\Bigg>_{\text{connected}} \notag \\
	&\,\,=
		\paren{- 4 \lambda_0 g_0^2 \mu^{4\vep}}
		\int_p \frac{e^{+i p\cdot\paren{x-y} -\paren{t+s}p^2}}{ \paren{p^2 + m_0^2}^2 }
		\int_\ell \frac{1}{ \ell^2 + m_0^2 },
	\label{eq:Psi-2PTform-b} \\
\left< \Psi\fn{t,x} \Psi^\dagger\fn{s,y} \right>_\text{c}
	&:=
		\Bigg<
		\wt{\Psi}\fn{t,x} \wt{\Psi}^\dagger\fn{s,y}
		\frac{1}{2}
		\br{
			\paren{-1}
			\int \df^d X \sqbr{  i A_\alpha \paren{ \Phi^\dagger \p_\alpha \Phi - \paren{\p_\alpha \Phi^\dagger} \Phi  }_{\fn{X}}  }
		} \notag \\
	&\,\,\quad \times
		\br{
			\paren{-1}
			\int \df^d Y \sqbr{  i A_\beta \paren{ \Phi^\dagger \p_\beta \Phi - \paren{\p_\beta \Phi^\dagger} \Phi  }_{\fn{Y}}  }
		}
		\Bigg>_{\text{connected}} \notag \\
	&\,\,=
		{\paren{ g_0^4 \mu^{4\vep}}}
		\int_p \frac{e^{+i p\cdot\paren{x-y} -\paren{t+s}p^2}}{ \paren{p^2 + m_0^2}^2 }
		\int_\ell \frac{ \paren{\ell+p}^2 }{  \paren{\ell-p}^2 \paren{\ell^2+m_0^2}  },
	\label{eq:Psi-2PTform-c}
}
where a virtual mass for the gauge boson $\mu_A$ is introduced for the calculation of diagram (a).

The last {five} contributions, (B$'$) to {(F$'$)}, are obtained by parameter replacement and complex conjugation as
\al{
\left< \Psi\fn{t,x} \Psi^\dagger\fn{s,y} \right>_\text{B$'$}
	&=
		\paren{\left< \Psi\fn{s,y} \Psi^\dagger\fn{t,x} \right>_\text{B}}^\ast,&
\left< \Psi\fn{t,x} \Psi^\dagger\fn{s,y} \right>_\text{C$'$}
	&=
		\paren{\left< \Psi\fn{s,y} \Psi^\dagger\fn{t,x} \right>_\text{C}}^\ast, \notag \\
\left< \Psi\fn{t,x} \Psi^\dagger\fn{s,y} \right>_\text{D$'$}
	&=
		\paren{\left< \Psi\fn{s,y} \Psi^\dagger\fn{t,x} \right>_\text{D}}^\ast,&
\left< \Psi\fn{t,x} \Psi^\dagger\fn{s,y} \right>_\text{E$'$}
	&=
		\paren{\left< \Psi\fn{s,y} \Psi^\dagger\fn{t,x} \right>_\text{E}}^\ast, \notag \\
{\left< \Psi\fn{t,x} \Psi^\dagger\fn{s,y} \right>_\text{F$'$}}
	&=
		{\paren{\left< \Psi\fn{s,y} \Psi^\dagger\fn{t,x} \right>_\text{F}}^\ast.} &
	\label{eq:replacement-rule}
}
From the above forms, we recognize that no singularity emerges in the limit $y \to x$, and the expectation value of
the smeared version of the composite operator
$\widehat{\Psi}^\dagger\fn{x}\widehat{\Psi}\fn{x}$ is well defined.

Also, we need the counter term~(CT) from the bulk wave function renormalization of the flowed field,
which is easily estimated from Eqs.~\eqref{eq:Psi-2PTform-LO} and \eqref{eq:bulk-scalar-counterpart-formal} as
\al{
\left< \Psi\fn{t,x} \Psi^\dagger\fn{s,y} \right>_\text{CT}
	:=&\,
		\delta{\cal Z}_\Psi 
		\left< \wt{\Psi}\fn{t,x} \wt{\Psi}^\dagger\fn{s,y} \right>_0 \notag \\
	=&\,
		\delta{\cal Z}_\Psi 
		\paren{ g_0^2 \mu^{2\vep} }
		\int_p \frac{ e^{ip \cdot \paren{x-y} - \paren{t+s} p^2 } }{ p^2 + m_0^2 }.
		\label{eq:Psi-2PTform-CT}
}

\subsection{Evaluation of momentum integrals}

The smeared two-point functions are regular in the limit $y\to x$, and we can estimate them approximately well by use of
the saddle-point method when the flow times $s$ and $t$ are sufficiently large, since the solution to the gradient flow has a form such as $\int_p f(p) e^{-tp^2}$. 
We frequently utilize the saddle-point approximation formula,
\al{
\int_p f\fn{p} e^{- \alpha \paren{p - p_\ast}^2}
	\simeq
		\paren{ \frac{1}{4\pi \alpha} }^{d/2} f\fn{p_\ast},
	\label{eq:SPA-formula}
}
where this approximation works fine for a sufficiently large positive $\alpha$,
$f\fn{p}$ is a rational function of the $d$-dimensional momentum $p$,
and $p_\ast$ represents the position of the saddle point.

Here, UV divergences appear in the present calculations executed under the continuous spacetime limit due to intermediate loop diagrams.
{Also, we will observe the emergence of IR divergences, which may originate from saddle-point calculations around $p_\ast \simeq 0$.
They are considered artifacts due to our approximation with the saddle-point method.
As we will see, if we evaluate divergent parts without the approximation, IR divergences do not emerge, as expected.\footnote{It is understood that the IR divergence in Eq.~\eref{eq:Psi-2PTform-a} is cut off by $\mu_A$ as in the ordinary QED calculations, and we safely take the limit $\mu_A \to 0$ after the integral. 
}}
Details of the following calculations are available in Appendix~\ref{sec:Detail-HTPF}.

After the $\vep$ expansion around zero, we obrtain the form for the LO part,
\al{
\left< \Psi\fn{t,x} \Psi^\dagger\fn{s,x} \right>_\text{LO}
	=&
		\frac{g_0^2}{\paren{4\pi}^2} m_0^2 e^{m_0^2\paren{s+t}} \Gamma\fn{-1, m_0^2\paren{s+t}}
		+ {\cal O}\fn{\vep} \notag \\
	\xrightarrow[\text{large}\,s,\,t]{}&
		\frac{g_0^2}{\paren{4\pi}^2} \frac{1}{m_0^2 \paren{s+t}^2} + {\cal O}\fn{\vep},
		\label{saddle-point equation}
}
where we took the first term of the series expansion of $e^{m_0^2\paren{s+t}} \Gamma\fn{-1, m_0^2\paren{s+t}}$
around infinity.
This form is suitable for our current strategy, where the saddle-point approximation is adapted for the one-loop diagrams, assuming that $s$ and $t$ are sufficiently large.
Also, we derive the $\vep$-expanded form of the counter term in the same way,
\al{
\left< \Psi\fn{t,x} \Psi^\dagger\fn{s,x} \right>_\text{CT}
	\xrightarrow[\text{large}\,s,\,t]{}
		\left.{\cal Z}_\Psi 
		\frac{g_0^2}{\paren{4\pi}^2} \frac{1}{m_0^2 \paren{s+t}^2}\right|_\text{CT}+ {\cal O}\fn{\vep}.
}

For the contributions (A) to (E), results are shown as follows:
\al{
&\left< \Psi\fn{t,x} \Psi^\dagger\fn{s,x} \right>_\text{A}
	\xrightarrow[\text{s.p.}]{}
		\frac{g_0^4}{\paren{4\pi}^4\paren{s+t}^2 m_0^2}
		\frac{1}{4} \sqbr{-1-\log\fn{a}} \notag \\
	&\quad -
		\frac{9 g_0^4}{\paren{4\pi}^4\paren{s+t}^2 m_0^2}
		\Bigg\{
		\frac{1}{\vep} + 1 + \log\sqbr{2m_0^2\fn{s+t}} + \log\sqbr{ \frac{32\pi^2 \mu^4 \fn{s+t}}{m_0^2} }
		+ e^{2m_0^2\fn{s+t}} G^{2,0}_{1,2}\fn{ {}^{1}_{0,0} \Big| \, 2 m_0^2 \fn{s+t} }
		\Bigg\} \notag \\
	&\quad +
		\frac{9 g_0^4}{\paren{4\pi}^4 {s}^2 m_0^2}
		\Bigg\{
		\frac{1}{\vep} + 1 + \log\sqbr{2m_0^2\fn{s+2t}} + \log\sqbr{ \frac{32\pi^2 \mu^4 {s}}{m_0^2} }
		+ e^{2m_0^2\fn{s+2t}} G^{2,0}_{1,2}\fn{ {}^{1}_{0,0} \Big| \, 2 m_0^2 \fn{s+2t} }
		\Bigg\} \notag \\
	&\quad +
		\frac{9 g_0^4}{\paren{4\pi}^4 {t}^2 m_0^2}
		\Bigg\{
		\frac{1}{\vep} + 1 + \log\sqbr{2m_0^2\fn{2s+t}} + \log\sqbr{ \frac{32\pi^2 \mu^4 {t}}{m_0^2} }
		+ e^{2m_0^2\fn{2s+t}} G^{2,0}_{1,2}\fn{ {}^{1}_{0,0} \Big| \, 2 m_0^2 \fn{2s+t} }
		\Bigg\} + {\cal O}\fn{\vep}, 
		\label{part A}
		\\
&\left< \Psi\fn{t,x} \Psi^\dagger\fn{s,x} \right>_\text{B}
	\xrightarrow[\text{s.p.}]{}
		-\frac{g_0^4}{\paren{4\pi}^4\paren{s+t}^2 m_0^2}
		\br{
			\frac{2}{\vep} + 1 + 2 \log\sqbr{32\pi^2 t \paren{s+t} \mu^4}
		} + {\cal O}\fn{\vep}, \\
&\left< \Psi\fn{t,x} \Psi^\dagger\fn{s,x} \right>_\text{C}
	\xrightarrow[\text{s.p.}]{}
		\frac{g_0^4}{\paren{4\pi}^4\paren{s+t}^2 m_0^2}
		\frac{1}{2}
		\br{  \frac{1}{\vep} - \gamma + 1 + \log\sqbr{ \frac{16\pi^2\fn{s+t}\mu^4}{m_0^2} }  } \notag \\
	&\quad +
		\frac{6 g_0^4}{\paren{4\pi}^4{t}^2 m_0^2}
		e^{ \frac{1}{2} m_0^2 \paren{2s+t} }
		\sqbr{  \Gamma\fn{0, \frac{1}{2} m_0^2 \paren{2s+t}} - e^{ \frac{3}{2}m_0^2\paren{2s+t} } \Gamma\fn{0, 2 m_0^2 \paren{2s+t}} }
		+ {\cal O}\fn{\vep}, \\
&\left< \Psi\fn{t,x} \Psi^\dagger\fn{s,x} \right>_\text{D}
	\xrightarrow[\text{s.p.}]{}
		\frac{g_0^4}{\paren{4\pi}^4\paren{s+t}^2 m_0^2}
		\frac{1}{4}
		\fn{ \frac{1}{\vep} - \gamma + \log\sqbr{  \frac{16\pi^2\fn{s+t}\mu^4}{\mu_\text{IR}^2}  } } \notag \\
	&\quad -
		\frac{g_0^4}{\paren{4\pi}^4{t}^2 m_0^2}
		\frac{9}{2}
		\br{
			\frac{1}{\vep} + 1 + \log\sqbr{16\pi^2t\fn{2s+t}\mu^4  } 
			+ e^{\frac{1}{2}m_0^2\fn{2s+t}} G^{2,0}_{1,2}\fn{ {}^{1}_{0,0} \Big| \, \frac{1}{2} m_0^2 \fn{2s+t} }
		} \notag \\
	&\quad -
		\frac{g_0^4(s+3 t)^2 (s+5 t)^2}{\paren{4\pi}^4 8 t (s+t)^5m_0^2 }
		\br{
			 \frac{1}{\vep} + 1+ \log\left[ 32\pi^2 t(s+t) \mu^4\right] 
				+ e^{ \frac{m_0^2 (s+t) (s+3 t)}{2 t}  } G^{2,0}_{1,2}\fn{ {}^{1}_{0,0} \Big| \, \frac{m_0^2 (s+t) (s+3 t)}{2 t} } 
		}\notag\\
		 &\quad+ {\cal O}\fn{\vep}, \\
&\left< \Psi\fn{t,x} \Psi^\dagger\fn{s,x} \right>_\text{E}
	\xrightarrow[\text{s.p.}]{}
		- \frac{g_0^4}{\paren{4\pi}^4\paren{s+t}^2 m_0^2}
		\frac{1}{4}
		\br{ \frac{1}{\vep} - \gamma +1 + \log\sqbr{  \frac{16\pi^2\paren{s+t}\mu^4}{m_0^2}  } } \notag \\
	&\quad -
		\frac{g_0^4}{\paren{4\pi}^4{t}^2 m_0^2}
		\frac{3}{2}
		e^{ \frac{1}{2} m_0^2 \paren{2s+t} }
		\sqbr{  \Gamma\fn{0, \frac{1}{2} m_0^2 \paren{2s+t}}
			 -  \, e^{ \frac{3}{2}m_0^2\paren{2s+t} } \Gamma\fn{0, 2 m_0^2 \paren{2s+t}} }
		+ {\cal O}\fn{\vep},
		\label{part E}
}
where ``s.p.'' indicates adoption of the saddle-point approximation.
Here, $a$ is a dimensionless regulator for an infrared divergence, and $\mu_\text{IR}$ is a virtual mass for the massless gauge boson to regularize an infrared divergence.
$\Gamma\fn{a,z}$ is the incomplete gamma function defined by
\al{
\Gamma\fn{a,z} := \int_z^\infty t^{a-1} e^{-t} \df t,
}
where $\Gamma\fn{a,0}$ is reduced to the Euler gamma function $\Gamma\fn{a}$
and $\gamma$ is the Euler--Mascheroni constant.
$G^{2,0}_{1,2}\fn{ {}^{1}_{0,0} \Big| \, z }$ is the Meijer G-function under the designated arguments.
The definition of the Meijer G-function is
\al{
G^{m,n}_{p,q} \paren{ {}^{a_1,...,a_p}_{b_1,...,b_q} \Big| z}
	&:=
		\frac{1}{2\pi i}
		\int_{\gamma_L}
		\frac
			{ \Pi_{j=1}^m \Gamma\fn{b_j+s} \Pi_{j=1}^n \Gamma\fn{1 - a_j - s} }
			{ \Pi_{j=n+1}^p \Gamma\fn{a_j+s} \Pi_{j=m+1}^q \Gamma\fn{1 - a_j - s} }
		z^{-s} \df s,
}
where $m$, $n$, $p$, and $q$ are integers obeying $0 \leq n \leq p$ and $0 \leq m \leq q	$, and
the contour $\gamma_L$ lies between the poles of $\Gamma\fn{1-a_j-s}$ and the poles of $\Gamma\fn{b_i+s}$~\cite{MeijerG-WolframMathworld}.
The four corresponding results of (B$'$) to (E$'$) are easily obtainable using the replacement rule in Eq.~\eqref{eq:replacement-rule}.

{We can show that the bulk tadpole contribution (F) vanishes without any approximations:
\al{
\left< \Psi\fn{t,x} \Psi^\dagger\fn{s,x} \right>_\text{F} = 0.
}
This means that only the 1PI configurations provide nonzero contributions to the two-point function.}

For the configurations (a) to (c), we reach the $\vep$-expanded forms, 
\al{
\left< \Psi\fn{t,x} \Psi^\dagger\fn{s,x} \right>_\text{a}
	&=
		0, 
		\label{part a}\\
\left< \Psi\fn{t,x} \Psi^\dagger\fn{s,x} \right>_\text{b}
	&=
		4 \lambda_0 g_0^2 \frac{m_0^2}{\paren{4\pi}^4}
		\Bigg\{
			\frac{1}{\vep} \sqbr{  -1 + e^{m_0^2\paren{s+t}} \paren{1 + m_0^2\paren{s+t}} E_1\fn{ m_0^2\paren{s+t} } } \notag \\
	&\quad \quad +
		\paren{ 2-\gamma + \log\sqbr{\frac{16\pi^2\paren{s+t}\mu^4}{m_0^2}} } 
			\sqbr{  -1 + e^{m_0^2\paren{s+t}} \paren{1 + m_0^2\paren{s+t}} E_1\fn{ m_0^2\paren{s+t} } } \notag \\
	&\quad \quad -
		e^{m_0^2\paren{s+t}} E_1\fn{ m_0^2\paren{s+t} } +
		e^{m_0^2\paren{s+t}} \paren{1 + m_0^2\paren{s+t}}
		G^{3,0}_{2,3}\fn{ {}^{1,1}_{0,0,0} \Big| \, m_0^2\paren{s+t} }
	\Bigg\} + {\cal O}\fn{\vep}, \\
\left< \Psi\fn{t,x} \Psi^\dagger\fn{s,x} \right>_\text{c}
	&\xrightarrow[\text{s.p.}]{}
		\frac{g_0^4}{\paren{4\pi}^4\paren{s+t}^2 m_0^2}
		\br{  -\frac{1}{\vep} + \gamma -1 + \log\sqbr{  \frac{m_0^2}{16\pi^2\paren{s+t}\mu^4}  }  } + {\cal O}\fn{\vep},
		\label{part c}
}
where $E_n\fn{z}$ represents the exponential integral function defined by
\al{
E_n\fn{z}
	&:=
		\int_1^{\infty} \frac{e^{-z t}}{t^n} \df t,
}
and $G^{3,0}_{2,3}\fn{ {}^{1,1}_{0,0,0} \Big| \, z }$ is another specific case of the Meijer G-function under the designated arguments.

\subsubsection{Divergent part in the saddle-point approximation}


In each final expression written below in the renormalized couplings, we ignore higher-order parts in multiplicative computations:
\al{
\left.\left< \Psi\fn{t,x} \Psi^\dagger\fn{s,x} \right>_\text{A}^\text{(s.p.)}\right|_\text{div}
	&={ \frac{g^4}{(4\pi)^4   m^2}\left(\frac{1}{t^2}+\frac{1}{s^2}-\frac{1}{(t+s)^2}\right)\frac{9}{\varepsilon}+\frac{g^4}{\paren{4\pi}^4\paren{t+s}^2 m^2}\frac{1}{4} \log\fn{a}}{,}\\
\left.\left< \Psi\fn{t,x} \Psi^\dagger\fn{s,x} \right>_\text{B}^\text{(s.p.)}\right|_\text{div}
	&=
		-\frac{g^4}{\paren{4\pi}^4\paren{t+s}^2 m^2}
			\frac{2}{\vep}, \\
\left.\left< \Psi\fn{t,x} \Psi^\dagger\fn{s,x} \right>_\text{C}^\text{(s.p.)}\right|_\text{div}
	&=
		\frac{g^4}{\paren{4\pi}^4\paren{t+s}^2 m^2}
		\frac{1}{2}
		\frac{1}{\vep}, \\
\left.\left< \Psi\fn{t,x} \Psi^\dagger\fn{s,x} \right>_\text{D}^\text{(s.p.)}\right|_\text{div}
	&=
	{\frac{	{g^4}}{(4\pi)^4 m^2}\left\{ \frac{1}{4}\frac{1}{(t+s)^2}-\frac{9}{2}\frac{1}{t^2}  +  \frac{(s+3 t)^2 (s+5 t)^2}{8 t (s+t)^5}\right\}\frac{1}{\varepsilon}}\notag\\
	&\quad{	+\frac{g^4}{\paren{4\pi}^4\paren{t+s}^2 m^2}
	\sqbr{
			 \frac{1}{4} \log\fn{ \frac{16\pi^2\paren{t+s}\mu^4}{\mu_\text{IR}^2} }
		}{,}}  \\
\left.\left< \Psi\fn{t,x} \Psi^\dagger\fn{s,x} \right>_\text{E}^\text{(s.p.)}\right|_\text{div}
	&=
		- \frac{g^4}{\paren{4\pi}^4\paren{t+s}^2 m^2}
		\frac{1}{4}
		\frac{1}{\vep}, \\
{\left.\left< \Psi\fn{t,x} \Psi^\dagger\fn{s,x} \right>_\text{F}\right|_\text{UV-div}} &= 0, \\
\left.\left< \Psi\fn{t,x} \Psi^\dagger\fn{s,x} \right>_\text{a}\right|_\text{div}
	&=
		0, \\
\left.\left< \Psi\fn{t,x} \Psi^\dagger\fn{s,x} \right>_\text{b}\right|_\text{div}
	&=
		4 \lambda g^2 \frac{m^2}{\paren{4\pi}^4}
		\frac{1}{\vep}
		\br{  -1 + e^{m^2\paren{t+s}} \sqbr{1 + m^2\paren{t+s}} E_1\fn{ m^2\paren{t+s} } }, \\
\left.\left< \Psi\fn{t,x} \Psi^\dagger\fn{s,x} \right>_\text{c}^\text{(s.p.)}\right|_\text{div}
	&=
		-
		\frac{g^4}{\paren{4\pi}^4\paren{t+s}^2 m^2}
		\frac{1}{\vep}, \\
\left.\left< \Psi\fn{t,x} \Psi^\dagger\fn{s,x} \right>_\text{LO}^\text{(s.p.)}\right|_\text{div}
	&=
		\frac{g_0^2}{\paren{4\pi}^2} \frac{1}{m_0^2 \paren{t+s}^2}, \notag \\
	&=
		\frac{1}{\paren{4\pi}^2 \paren{t+s}^2 m^2}
		\sqbr{
			\frac{g^4}{3 \paren{4\pi}^2} \frac{1}{\vep} - \frac{1}{\paren{4\pi}^2} \paren{4g^2\lambda-3g^4} \frac{1}{\vep}
		},
		\\
\left.\left< \Psi\fn{t,x} \Psi^\dagger\fn{s,x} \right>_\text{CT}^\text{(s.p.)}\right|_\text{div}
	&=
		\delta {\cal Z}_\Psi 
		\frac{g^2}{\paren{4\pi}^2} \frac{1}{m^2 \paren{t+s}^2},
}
where the first and second terms of the result for the LO part originate from $g_0^2$ and $m_0^2$, respectively.
Here, we straightforwardly determine the condition for $\delta {\cal Z}_\Psi$ to eliminate both the UV and IR divergences listed above as
{
\begin{align}
\delta {\cal Z}_\Psi
	=&
		- \frac{g^2}{384 \pi^2} \frac{1}{\vep}
			\frac{108 \, s^6 + 435 \, s^5t + 1244 \, s^4 t^2 + 682 \, s^3 t^3 + 1244 \, s^2 t^4 + 
				435 \, s t^5 + 108 \, t^6}
			{s^2 t^2 \paren{t+s}^2} \notag \\
	&
		+ \frac{\lambda}{4 \pi^2} \frac{1}{\vep} \sqbr{ 1 + m^4 \paren{t+s}^2 }
		- \frac{\lambda}{4 \pi^2} \frac{1}{\vep}
			m^4 \paren{t+s}^2 e^{m^2\paren{t+s}} \sqbr{1 + m^2\paren{t+s}} E_1\fn{ m^2\paren{t+s} } \notag \\
	&
		- \frac{g^2}{64 \pi^2} \br{ \log\fn{a} + 2 \log\sqbr{ \frac{16 \pi^2 \paren{t+s} \mu^4}{\mu_\tx{IR}^2} } }.
	\label{eq:deltaZ_Psi_value}
\end{align}
}

\subsubsection{UV-divergent part without approximation}

On the other hand, we can identify the UV poles due to the integrals of the loop momentum $\ell$ before the inverse Fourier transforms
after taking the safe limit $y\to x$ as follows:
\al{
\left.\left< \Psi\fn{t,x} \Psi^\dagger\fn{s,x} \right>_\text{A}\right|_\text{UV-div}
	&=
		0, \\
\left.\left< \Psi\fn{t,x} \Psi^\dagger\fn{s,x} \right>_\text{B}\right|_\text{UV-div} &=
\left.\left< \Psi\fn{t,x} \Psi^\dagger\fn{s,x} \right>_\text{B$'$}\right|_\text{UV-div} 
		=
		\int_p \frac{e^{ -\paren{t+s}p^2}}{p^2 + m^2}
		\paren{ \frac{{- 2} g^4}{\paren{4\pi}^2} \frac{1}{\vep} }, \\
\left.\left< \Psi\fn{t,x} \Psi^\dagger\fn{s,x} \right>_\text{C}\right|_\text{UV-div} &=
\left.\left< \Psi\fn{t,x} \Psi^\dagger\fn{s,x} \right>_\text{C$'$}\right|_\text{UV-div}
		=
		\int_p \frac{e^{ -\paren{t+s}p^2}}{p^2 + m^2}
		\paren{ \frac{{g^4}}{2\paren{4\pi}^2} \frac{1}{\vep} }, \\
\left.\left< \Psi\fn{t,x} \Psi^\dagger\fn{s,x} \right>_\text{D}\right|_\text{UV-div} &=
\left.\left< \Psi\fn{t,x} \Psi^\dagger\fn{s,x} \right>_\text{D$'$}\right|_\text{UV-div} 
		=
		\int_p \frac{e^{ -\paren{t+s}p^2}}{p^2 + m^2}
		\paren{ \frac{{{+}g^4}}{4\paren{4\pi}^2} \frac{1}{\vep} }, \\
\left.\left< \Psi\fn{t,x} \Psi^\dagger\fn{s,x} \right>_\text{E}\right|_\text{UV-div} &=
\left.\left< \Psi\fn{t,x} \Psi^\dagger\fn{s,x} \right>_\text{E$'$}\right|_\text{UV-div}
		=
		\int_p \frac{e^{ -\paren{t+s}p^2}}{p^2 + m^2}
		\paren{   \frac{- g^4}{4 \paren{4\pi}^2}  \frac{1}{\vep}  }, \\
{\left.\left< \Psi\fn{t,x} \Psi^\dagger\fn{s,x} \right>_\text{F}\right|_\text{UV-div}} &= 0, \\
\left.\left< \Psi\fn{t,x} \Psi^\dagger\fn{s,x} \right>_\text{a}\right|_\text{UV-div}
	&=
		0,
		\label{eq:UV-div-a} \\
\left.\left< \Psi\fn{t,x} \Psi^\dagger\fn{s,x} \right>_\text{b}\right|_\text{UV-div}
	&=		
		\int_p \frac{e^{ -\paren{t+s}p^2}}{ \paren{p^2 + m^2}^2 }
		\paren{   \frac{ {4 \lambda g^2}m^2}{\paren{4\pi}^2} \frac{1}{\vep}   }, \\
\left.\left< \Psi\fn{t,x} \Psi^\dagger\fn{s,x} \right>_\text{c}\right|_\text{UV-div}
	&=		
		\int_p \frac{e^{ -\paren{t+s}p^2}}{ \paren{p^2 + m^2}^2 }
		\br{   \frac{ {g^4} \sqbr{ 2\paren{p^2 + m^2} -3m^2 } }{\paren{4\pi}^2} \frac{1}{\vep}   }, \\
\left.\left< \Psi\fn{t,x} \Psi^\dagger\fn{s,x} \right>_\text{LO}\right|_\text{UV-div}
	&=	
		\paren{ g_0^2 \mu^{2\vep} }
		\int_p \frac{e^{ -\paren{t+s}p^2}}{p^2 + m_0^2} \\
	&= 
		\int_p \frac{e^{ -\paren{t+s}p^2}}{p^2 + m^2}
		\sqbr{
			\frac{g^4}{3 \paren{4\pi}^2} \frac{1}{\vep} - \frac{1}{p^2+m^2} \frac{m^2}{\paren{4\pi}^2} \paren{4g^2\lambda-3g^4} \frac{1}{\vep}
		}, \\
\left.\left< \Psi\fn{t,x} \Psi^\dagger\fn{s,x} \right>_\text{CT}\right|_\text{UV-div}
	&=
		\paren{\delta {\cal Z}_\Psi g^2}
		\int_p \frac{e^{ -\paren{t+s}p^2}}{p^2 + m^2},
}
{where we ignore higher-order parts in multiplicative computations in each expression written in the renormalized couplings.}

The total of the UV-divergent parts shown above takes the form
\al{
\left.\left< \Psi\fn{t,x} \Psi^\dagger\fn{s,x} \right>_\text{total}\right|_\text{UV-div}
	&=
		\paren{ -\frac{2}{3} \frac{1}{\vep} \frac{g^4}{{(4\pi)^2}} + \delta {\cal Z}_\Psi g^2} \, 
		\int_p \frac{e^{ -\paren{t+s}p^2}}{p^2 + m^2},
}
where no divergence associated with $m$ or $\lambda$ remains, as expected.
We can remove the remaining divergence associated with $g$ by taking the bulk counterpart as
\al{
\delta {\cal Z}_\Psi = {\frac{g^2}{(4\pi)^2} \frac{2}{3} \frac{1}{\vep}.}
\label{wave_renormalize}
}
{Note that we do not adopt any approximations to identify the divergences, where only UV divergences emerge.
We confirmed that the IR divergences emerging in the previous calculation with the saddle-point approximation are artifacts.}

In other words, we can take the two-point function of the flowed Higgs fields as finite when we choose the flow field's extra wave function as \eref{wave_renormalize}. This result shows that the gradient flow indeed works in the $U(1)$ Higgs model at the one-loop level, {as shown for the non-Abelian gauge theory with a fermion as a matter field in Ref.~\cite{Luscher:2013cpa}.
After subtracting the divergent part, we concentrate on the finite part of the Higgs two-point function.
}

\subsection{Asymptotic form in large flow time}

Here, we consider the series expansion around infinity for $s=t$, after the coupling renormalizations and
the removal of the divergences by taking the bulk counter term $\delta {\cal Z}_\Psi$ suitably as shown in Eq.~\eqref{eq:deltaZ_Psi_value}: For a large $s$ corresponding to an IR region,
\al{
\left< \Psi\fn{s,x} \Psi^\dagger\fn{s,x} \right>_\text{LO}^\text{(s.p.)}
	&=
		\frac{g^2}{64 m^2 \pi^2 s^2} + {\cal O}\fn{\frac{1}{s^3}}, \\
\left< \Psi\fn{s,x} \Psi^\dagger\fn{s,x} \right>_\text{A}^\text{(s.p.)}
	&=
		- \frac{g^4}{4096 \pi^4 m^2 s^2}
		\Bigg[
			-251 + 36 \log\fn{4 m^2 s} - 288 \log\fn{6 m^2 s} \notag \\
	&\qquad\qquad
		-288 \log\fn{  \frac{ 32\pi^2 s \mu^4 }{m^2}  }
		+36 \log\fn{  \frac{ 64\pi^2 s \mu^4 }{m^2}  }
		\Bigg]  + {\cal O}\fn{\frac{1}{s^3}}, \\
\left< \Psi\fn{s,x} \Psi^\dagger\fn{s,x} \right>_\text{B}^\text{(s.p.)}
	&=
		- \frac{g^4}{1024 \pi^4 m^2 s^2}
		\Bigg[
			1 + 2 \log\fn{ 64 \pi^2 s^2 \mu^4 }
		\Bigg] + {\cal O}\fn{\frac{1}{s^3}}, \\
\left< \Psi\fn{s,x} \Psi^\dagger\fn{s,x} \right>_\text{C}^\text{(s.p.)}
	&=
		- \frac{g^4}{2048 \pi^4 m^2 s^2}
		\Bigg[
			-1 +\gamma - \log\fn{ \frac{32 \pi^2 s \mu^4}{m^2} }
		\Bigg] + {\cal O}\fn{\frac{1}{s^3}}, \\
\left< \Psi\fn{s,x} \Psi^\dagger\fn{s,x} \right>_\text{D}^\text{(s.p.)}
	&=
		- \frac{g^4}{4096 \pi^4 m^2 s^2}
		\Bigg[
			36 + \gamma + 72 \log\fn{\frac{3 m^2 s}{2}} - 36 \log\fn{4 m^2 s} \notag \\
	&\qquad\qquad
		-36 \log\fn{  \frac{ 16\pi^2 s \mu^4 }{m^2}  }
		+72 \log\fn{  \frac{ 32\pi^2 s \mu^4 }{m^2} }
		\Bigg]  + {\cal O}\fn{\frac{1}{s^3}}, \\
\left< \Psi\fn{s,x} \Psi^\dagger\fn{s,x} \right>_\text{E}^\text{(s.p.)}
	&=
		\frac{g^4}{4096 \pi^4 m^2 s^2}
		\Bigg[
			-1 + \gamma - \log\fn{ \frac{32 \pi^2 s \mu^4}{m^2} }
		\Bigg] + {\cal O}\fn{\frac{1}{s^3}}, \\
{\left< \Psi\fn{s,x} \Psi^\dagger\fn{s,x} \right>_\text{F}}
	&= {\left< \Psi\fn{s,x} \Psi^\dagger\fn{s,x} \right>_\text{F'} =
		0}, \\
\left< \Psi\fn{s,x} \Psi^\dagger\fn{s,x} \right>_\text{a}
	&=
		0, \\
\left< \Psi\fn{s,x} \Psi^\dagger\fn{s,x} \right>_\text{b}
	&=
		- \frac{\lambda g^2}{256 \pi^4 m^2 s^2}
		\Bigg[
			-1 +\gamma - \log\fn{ \frac{32 \pi^2 s \mu^4}{m^2} }
		\Bigg] + {\cal O}\fn{\frac{1}{s^3}}, \\
\left< \Psi\fn{s,x} \Psi^\dagger\fn{s,x} \right>_\text{c}^\text{(s.p.)}
	&=
		\frac{g^4}{1024 \pi^4 m^2 s^2}
		\Bigg[
			-1 +\gamma - \log\fn{ \frac{32 \pi^2 s \mu^4}{m^2} }
		\Bigg] + {\cal O}\fn{\frac{1}{s^3}},
}
with
\al{
\left< \Psi\fn{s,x} \Psi^\dagger\fn{s,x} \right>_\text{B$'$}^\text{(s.p.)} &= \left< \Psi\fn{s,x} \Psi^\dagger\fn{s,x} \right>_\text{B}^\text{(s.p.)},&
\left< \Psi\fn{s,x} \Psi^\dagger\fn{s,x} \right>_\text{C$'$}^\text{(s.p.)} &= \left< \Psi\fn{s,x} \Psi^\dagger\fn{s,x} \right>_\text{C}^\text{(s.p.)},& \notag \\
\left< \Psi\fn{s,x} \Psi^\dagger\fn{s,x} \right>_\text{D$'$}^\text{(s.p.)} &= \left< \Psi\fn{s,x} \Psi^\dagger\fn{s,x} \right>_\text{D}^\text{(s.p.)},&
\left< \Psi\fn{s,x} \Psi^\dagger\fn{s,x} \right>_\text{E$'$}^\text{(s.p.)} &= \left< \Psi\fn{s,x} \Psi^\dagger\fn{s,x} \right>_\text{E}^\text{(s.p.)}.&
}
Therefore at both levels of LO and the one-loop, we can conclude that
\al{
\left< \Psi\fn{s,x} \Psi^\dagger\fn{s,x} \right>^\text{(s.p.)}
	\xrightarrow[{s\to \infty}]{}
		0,
}
in the saddle-point approximation.
This result is just as expected for the spontaneous gauge symmetry-breaking order parameters in the symmetric phase.

\section{Summary and discussion}\label{5}

We have proposed the flowed order parameter for spontaneous gauge symmetry breaking: the expectation value of the manifestly gauge-invariant composite operator at the same spacetime point, $\wh\Phi^\dagger\fn{x}\wh\Phi\fn{x}$, smeared towards the flow-time direction $t$ under the gradient flow.
The deformed configurations $\Psi\fn{t,x}$ and $\Psi^\dagger\fn{s,y}$ after progressions of the flow times $t$ and $s$ have the good property:
the limit $y \to x$ can be taken without an extra UV divergence in its two-point function $\left< \Psi\fn{t,x} \Psi^\dagger\fn{s,y} \right>$ as long as  $t,s > 0$.

Here, $\left< \Psi\fn{t,x} \Psi^\dagger\fn{s,x} \right>=\left< \Psi^\dagger\fn{s,x} \Psi\fn{t,x} \right>$ is interpreted as a well-defined version of $\bigl\langle \wh\Phi^\dagger\fn{x}\wh\Phi\fn{x}\bigr\rangle$.
The diffusion in the $t$ direction through the gradient flow can be interpreted as decreasing the physical reference energy.
Therefore, under the limit $t=s \to \infty$, $\left< \Psi\fn{s,x} \Psi^\dagger\fn{s,x} \right>$ would be related to $\bigl\langle \wh\Phi^\dagger\fn{x}\wh\Phi\fn{x} \bigr\rangle$ that is evaluated with the ground state.
Therefore, the binary information of whether $\lim_{s\to \infty} \left< \Psi\fn{s,x} \Psi^\dagger\fn{s,x} \right>$ takes zero or a nonzero value might be used for determining the phase.

As a first step, we compute the flowed order parameter in the Abelian Higgs model with a positive mass-squared parameter in the continuum theory at the one-loop level. With the help of the saddle-point approximation, we have derived the asymptotic analytic form of $\left< \Psi\fn{s,x} \Psi^\dagger\fn{s,x} \right>$ for large $s$.

We have checked the following:
(i) the limit $\left< \Psi\fn{s,x} \Psi^\dagger\fn{s,x} \right> \to0$ for $s\to\infty$, which consistently implies that the theory is still in the symmetric phase after taking into account one-loop radiative corrections for a finite positive mass-squared;
(ii) the UV finiteness of the $U(1)$ gauge-boson two-point function at the one-loop level, which is a concrete confirmation of the magnificent property for gradient-flowed gauge fields, firstly discussed in Refs.~\cite{Luscher:2010iy,Luscher:2011bx}; and
(iii) the UV finiteness of the Higgs boson two-point function under a wave-function renormalization that subtracts the UV poles originating from loop integrals~\cite{Luscher:2013cpa}.

This paper is a first step in a new direction, determining the phase of a physical system by the flowed order parameter, namely the flowed bilinear of the Higgs at the same spacetime point. It can shed new light on identifying the phase of a physical system. Here, we have relied on perturbation under gauge fixing and have adopted the saddle-point approximation to derive analytic forms. Generalization to non-Abelian gauge theories and investigations in the lattice gauge theory, which does not require gauge fixing, will be important next steps.

Even within the simplest Abelian Higgs scenario, various aspects await further clarification:
\begin{itemize}
\item
The standard perturbative method for determining the phase of the Abelian Higgs model is to investigate the vacua of the Coleman--Weinberg effective potential~\cite{Coleman:1973jx,Weinberg:1973am,Weinberg:2015}, which tells us that a dynamical gauge symmetry breaking occurs if the squared Higgs mass is zero or takes a sufficiently small positive value.
This region of the parameter space is beyond the scope of our current analysis relying on the saddle-point approximation under the assumption that the flow times $t$ and $s$ are much greater than other dimensional parameters such as $s,t\gg m^{-2}$, that is, our computation relies on the saddle-point method~\eqref{saddle-point equation} for the integral of the form
$
\int_p \frac{e^{ -\paren{t+s}p^2}}{p^2 + m^2}g\fn{p},
$
where $g\fn{p}$ is a rational function.
It would be interesting to investigate the massless or nearly massless region, which might require a numerical calculation.
\item
As is widely known, if the squared Higgs mass parameter is negative, the gauge symmetry is spontaneously broken down through the Higgs mechanism. Investigating the general properties of the gradient flow for the gauge theory in the broken phase is a theoretically important task.
\item
The two-point function of the smeared fields at the same spacetime point is well defined.
However, the deformation through the gradient flow might modify some of the properties of the original fields.
Our result in the focused parameter region appears consistent, and further theoretical investigation would be worthwhile.
\end{itemize}

We expect that the theory space at $t=0$ has one-to-one correspondence to that at $t>0$.
For a nonzero $t$, the finite parts in Eqs.~\eqref{part A}--\eqref{part E} and \eqref{part a}--\eqref{part c} are nonzero even for a large positive mass-squared at finite $t$.
However, we should remember that the physical value of the flowed order parameter for the finite~$t$ can be defined only after we fix a renormalization condition for the theory at $t$, which will be investigated in future work.
This might lead to a new relationship between the gradient flow and the renormalization group as suggested in Refs.~\cite{Sint:2014pip,Makino:2018rys,Abe:2018zdc,Carosso:2018bmz,Sonoda:2019ibh,Sonoda:2020vut,Miyakawa:2021hcx,Miyakawa:2021wus,Abe:2022smm,Sonoda:2022fmk,Miyakawa:2022qbz,Hasenfratz:2022wll} in a different context.

\bigskip

\section*{Acknowledgement}

We thank Tetsuo Hatsuda, Etsuko Ito, Daisuke Kadoh, and Naoya Ukita for useful discussion.
This work is in part supported by JSPS KAKENHI Grant Numbers 18K13546~(KK), 19H01899 (KO), and 21H01107 (KN, KO).

\appendix
\section*{Appendix}


\section{Details on the Higgs two-point function at one loop
\label{sec:Detail-HTPF}}

In this section, we provide details on the calculations of the Higgs two-point function at one loop, where both the divergent and finite parts are discussed.

We recall the following standard formulas for the $\vep$ expansion:
\al{
\frac{1}{1-\vep}
	&=
		1 + \vep + {\cal O}\fn{\vep^2}, \notag \\
A^\vep
	&=
		1 + \vep \log{A} + {\cal O}\fn{\vep^2}, \notag \\
\Gamma\fn{\vep}
	&=
		\frac{1}{\vep} - \gamma + {\cal O}\fn{\vep}, \notag \\
\Gamma\fn{\vep-1}
	&=
		-\frac{1}{\vep} + \gamma-1 + {\cal O}\fn{\vep},
}
where the variable $A$ is positive.
Also, we skip showing explicit forms for the contributions designated by the diagrams B$'$, C$'$, D$'$, and E$'$, since each part can be easily obtained from the corresponding result given below and the replacement rule in Eq.~\eqref{eq:replacement-rule}.

\subsection{Leading order}

For Eq.~\eqref{eq:Psi-2PTform-LO}, after taking the limit $y \to x$ safely, no approximation is necessary to perform the $p$ integral:
\al{
\left< \Psi\fn{t,x} \Psi^\dagger\fn{s,x} \right>_\text{LO}
	&=
		\paren{ g_0^2 \mu^{2\vep} }
		\int_p \frac{ e^{ - \paren{t+s} p^2 } }{ p^2 + m_0^2 } \notag \\
	&=
		\paren{ g_0^2 \mu^{2\vep} }
		\paren{ \frac{2}{\paren{4\pi}^{d/2} \Gamma\fn{d/2}} }
		\int_0^\infty \df p p^{d-1} \frac{ e^{ - \paren{t+s} p^2 } }{ p^2 + m_0^2 } \notag \\
	&=
		\paren{ g_0^2 \mu^{2\vep} }
		\paren{ \frac{2}{\paren{4\pi}^{d/2} \Gamma\fn{d/2}} }
		\frac{1}{2} e^{m_0^2 \paren{s+t}} m_0^{d-2} \Gamma\fn{\frac{d}{2}} \Gamma\fn{1-\frac{d}{2}, m_0^2 \paren{s+t}},
		\label{eq:Psi-LO2PT-exact}
}
where in the second and third lines, we integrated the isotropic angular part and the radial part of the $d$-dimensional $p$ integral, respectively.

Based on the property
\al{
\Gamma\fn{1-\frac{d}{2}, m_0^2 \paren{s+t}}
	&=
		\Gamma\fn{-1, m_0^2 \paren{s+t}} + {\cal O}\fn{\vep},
}
it is straightforward to obtain the $\vep$-expanded form,\footnote{
The following form for larger $t$ and $s$ is also obtained through the saddle-point method for the $p$ integral of
the second to last form of Eq.~\eqref{eq:Psi-LO2PT-exact}.
}
\al{
\left< \Psi\fn{t,x} \Psi^\dagger\fn{s,x} \right>_\text{LO}
	=& \,
		\frac{g_0^2}{\paren{4\pi}^2} m_0^2 e^{m_0^2\paren{s+t}} \Gamma\fn{-1, m_0^2\paren{s+t}}
		+ {\cal O}\fn{\vep} \notag \\
	\xrightarrow[\text{large}\,s,\,t]{}& \,
		\frac{g_0^2}{\paren{4\pi}^2} \frac{1}{m_0^2 \paren{s+t}^2} + {\cal O}\fn{\vep},
}
where we took the first term of the series expansion of $e^{m_0^2\paren{s+t}} \Gamma\fn{-1, m_0^2\paren{s+t}}$
around infinity,
\al{
e^{m_0^2\paren{s+t}} \Gamma\fn{-1, m_0^2\paren{s+t}}
	&=
		\paren{\frac{1}{m_0^2\paren{s+t}}}^2 + {\cal O}\sqbr{ \paren{\frac{1}{m_0^2\paren{s+t}}}^3 }.
}

\subsection{One-loop order}

\subsubsection{Diagram A}

In Eq.~\eqref{eq:Psi-2PTform-A}, after taking the limit $y \to x$ safely,
we can perform the integrals on the flow times exactly, where Eq.~\eqref{eq:Psi-2PTform-A} leads to
\al{
\left< \Psi\fn{t,x} \Psi^\dagger\fn{s,x} \right>_\text{A}
	&=
		\paren{g_0^4 \mu^{4\vep}}
		\int_{p,\ell} \frac{\paren{p+\ell}^2}{ \paren{p-\ell}^2 \paren{\ell^2 + m_0^2} }
		e^{-\paren{t+s}p^2} \notag \\
	&\quad
		\times
		\frac{1}{4 \paren{\ell^2 - \ell \cdot p}^2}
		\br{
			e^{-2\paren{s+t}\paren{\ell^2-\ell\cdot p}} - e^{-2s\paren{\ell^2-\ell\cdot p}}
			- e^{-2t\paren{\ell^2-\ell\cdot p}} + 1
		},
}
where we divide this form into the following four pieces:
\al{
\left< \Psi\fn{t,x} \Psi^\dagger\fn{s,x} \right>_\text{A-(i)}
	&:=
		+
		\paren{g_0^4 \mu^{4\vep}}
		\int_{p,\ell} \frac{\paren{p+\ell}^2}{ \paren{p-\ell}^2 \paren{\ell^2 + m_0^2} }
		e^{-\paren{t+s}p^2}
		\times
		\frac{1}{4 \paren{\ell^2 - \ell \cdot p}^2}
		e^{-2\paren{s+t}\paren{\ell^2-\ell\cdot p}}, 
		\label{eq:Psi-2PTform-A1} \\
\left< \Psi\fn{t,x} \Psi^\dagger\fn{s,x} \right>_\text{A-(ii)}
	&:=
		-
		\paren{g_0^4 \mu^{4\vep}}
		\int_{p,\ell} \frac{\paren{p+\ell}^2}{ \paren{p-\ell}^2 \paren{\ell^2 + m_0^2} }
		e^{-\paren{t+s}p^2}
		\times
		\frac{1}{4 \paren{\ell^2 - \ell \cdot p}^2}
		e^{-2s\paren{\ell^2-\ell\cdot p}}, 
		\label{eq:Psi-2PTform-A2} \\
\left< \Psi\fn{t,x} \Psi^\dagger\fn{s,x} \right>_\text{A-(iii)}
	&:=
		-
		\paren{g_0^4 \mu^{4\vep}}
		\int_{p,\ell} \frac{\paren{p+\ell}^2}{ \paren{p-\ell}^2 \paren{\ell^2 + m_0^2} }
		e^{-\paren{t+s}p^2}
		\times
		\frac{1}{4 \paren{\ell^2 - \ell \cdot p}^2}
		e^{-2t\paren{\ell^2-\ell\cdot p}}, 
		\label{eq:Psi-2PTform-A3} \\
\left< \Psi\fn{t,x} \Psi^\dagger\fn{s,x} \right>_\text{A-(iv)}
	&:=
		+
		\paren{g_0^4 \mu^{4\vep}}
		\int_{p,\ell} \frac{\paren{p+\ell}^2}{ \paren{p-\ell}^2 \paren{\ell^2 + m_0^2} }
		e^{-\paren{t+s}p^2}
		\times
		\frac{1}{4 \paren{\ell^2 - \ell \cdot p}^2}.
		\label{eq:Psi-2PTform-A4}
}

For Eq.~\eqref{eq:Psi-2PTform-A1}, the square completion of the exponent,
\al{
-\paren{t+s}p^2 -2\paren{s+t}\paren{\ell^2-\ell\cdot p}
	=
		-2\paren{s+t}\paren{\ell - \frac{p}{2}}^2
		-\frac{1}{2}\paren{s+t}p^2,
}
tells us the valid information required to make the saddle-point approximation for the integral of $\ell$.
The formula in Eq.~\eqref{eq:SPA-formula} with $\ell_\ast = p/2$ in Eq.~\eqref{eq:Psi-2PTform-A1} brings us to
\al{
\left< \Psi\fn{t,x} \Psi^\dagger\fn{s,x} \right>_\text{A-(i)}
	&\simeq
		\paren{g_0^4 \mu^{4\vep}}
		\paren{9\cdot 4^2}
		\paren{ \frac{1}{8\pi\paren{s+t}} }^{d/2}
		\int_p \frac{1}{\paren{p^2}^2 \paren{p^2 + 4m_0^2}}
		e^{ -\frac{1}{2} \paren{s+t} p^2 } \notag \\
	&=
		\paren{g_0^4 \mu^{4\vep}}
		\paren{9\cdot 4^2}
		\paren{ \frac{1}{8\pi\paren{s+t}} }^{d/2}
		\paren{ \frac{2}{ \paren{4\pi}^{d/2} \Gamma\fn{d/2} } } \notag \\
	&\quad
		\times
		\int_0^\infty \df p \, p^{d-1} \int_p \frac{1}{\paren{p^2}^2 \paren{p^2 + 4m_0^2}} e^{ -\frac{1}{2} \paren{s+t} p^2 } \notag \\
	&=
		\paren{g_0^4 \mu^{4\vep}}
		\paren{9\cdot 4^2}
		\paren{ \frac{1}{8\pi\paren{s+t}} }^{d/2}
		\paren{ \frac{2}{ \paren{4\pi}^{d/2} \Gamma\fn{d/2} } } \notag \\
	&\quad
		\times
		2^{-7+d} e^{2m_0^2\paren{s+t}} m_0^{-6+d}
		\Gamma\fn{-2+\frac{d}{2}} \Gamma\fn{ 3-\frac{d}{2}, 2m_0^2\paren{s+t} },
}
where in the second line we integrated the isotropic angular part of the $d$-dimensional momentum.
The evaluations of Eqs.~\eqref{eq:Psi-2PTform-A2} and \eqref{eq:Psi-2PTform-A3} can follow the same line, and their final forms are
\al{
\left< \Psi\fn{t,x} \Psi^\dagger\fn{s,x} \right>_\text{A-(ii)}
	&\simeq
		-
		\paren{g_0^4 \mu^{4\vep}}
		\paren{9\cdot 4^2}
		\paren{ \frac{1}{8\pi s} }^{d/2}
		\paren{ \frac{2}{ \paren{4\pi}^{d/2} \Gamma\fn{d/2} } } \notag \\
	&\quad
		\times
		2^{-7+d} e^{2m_0^2\paren{s+2t}} m_0^{-6+d}
		\Gamma\fn{-2+\frac{d}{2}} \Gamma\fn{ 3-\frac{d}{2}, 2m_0^2\paren{s+2t} }, \\
\left< \Psi\fn{t,x} \Psi^\dagger\fn{s,x} \right>_\text{A-(iii)}
	&\simeq
		-
		\paren{g_0^4 \mu^{4\vep}}
		\paren{9\cdot 4^2}
		\paren{ \frac{1}{8\pi t} }^{d/2}
		\paren{ \frac{2}{ \paren{4\pi}^{d/2} \Gamma\fn{d/2} } } \notag \\
	&\quad
		\times
		2^{-7+d} e^{2m_0^2\paren{2s+t}} m_0^{-6+d}
		\Gamma\fn{-2+\frac{d}{2}} \Gamma\fn{ 3-\frac{d}{2}, 2m_0^2\paren{2s+t} }.
}

For Eq.~\eqref{eq:Psi-2PTform-A4}, it is straightforward to apply the formula about the $p$ integral around the saddle point $p_\ast = 0$,
and we obtain the result
\al{
\left< \Psi\fn{t,x} \Psi^\dagger\fn{s,x} \right>_\text{A-(vi)}
	&\simeq
		\paren{g_0^4 \mu^{4\vep}}
		\paren{ \frac{1}{4\pi\paren{s+t}} }^{d/2}
		\int_\ell \frac{1}{4} \frac{1}{\paren{\ell^2}^2} \frac{1}{\ell^2 + m_0^2} \notag \\
	&=
		\paren{g_0^4 \mu^{4\vep}}
		\paren{ \frac{1}{4\pi\paren{s+t}} }^{d/2}
		\frac{1}{4} \frac{2!}{\paren{4\pi}^{d/2}} \frac{\Gamma\fn{3 - d/2}}{\Gamma\fn{3}} \int_0^1 \!\! \df x \int_0^{1-x} \!\!\!\! \df y
		\sqbr{ \frac{1}{ \paren{1-x-y} m_0^2 } }^{3-d/2},
		\label{eq:Psi-2PTform-A4p}
}
where we performed the $\ell$ integration with the help of the Feynman parameterization in a standard method.

The first three pieces are $\vep$-expanded straightforwardly, where the formulas are useful,
\al{
\frac{\Gamma\fn{-\vep}}{\Gamma\fn{2-\vep}}
	&=
		-\frac{1}{\vep} -1 -\vep + {\cal O}\fn{\vep^2}, 
		\label{eq:vep-formula-1} \\
\Gamma\fn{\vep+1, A}
	&=
		e^{-A} +
		\sqbr{
			e^{-A}  \log\fn{ A }
			+
			G^{2,0}_{1,2}\fn{ {}^{1}_{0,0} \Big| \, A }
		} \vep + {\cal O}\fn{\vep^2},
		\label{eq:vep-formula-2}
}
for a positive variable $A$.
The individual results are
\al{
&\left< \Psi\fn{t,x} \Psi^\dagger\fn{s,x} \right>_\text{A-(i)} \simeq \notag \\
	&
		\hspace{-7mm}
		-
		\frac{9 g_0^4}{\paren{4\pi}^4\paren{s+t}^2 m_0^2}
		\Bigg\{
		\frac{1}{\vep} + 1 + \log\sqbr{2m_0^2\fn{s+t}} + \log\sqbr{ \frac{32\pi^2 \mu^4 \fn{s+t}}{m_0^2} }
		+ e^{2m_0^2\fn{s+t}} G^{2,0}_{1,2}\fn{ {}^{1}_{0,0} | \, 2 m_0^2 \fn{s+t} }
		\Bigg\} + {\cal O}\fn{\vep}, \\
&\left< \Psi\fn{t,x} \Psi^\dagger\fn{s,x} \right>_\text{A-(ii)} \simeq \notag \\
	&
		+
		\frac{9 g_0^4}{\paren{4\pi}^4 {s}^2 m_0^2}
		\Bigg\{
		\frac{1}{\vep} + 1 + \log\sqbr{2m_0^2\fn{s+2t}} + \log\fn{ \frac{32\pi^2 \mu^4 {s}}{m_0^2} }
		+ e^{2m_0^2\fn{s+2t}} G^{2,0}_{1,2}\fn{ {}^{1}_{0,0} | \, 2 m_0^2 \fn{s+2t} }
		\Bigg\} + {\cal O}\fn{\vep}, \\
&\left< \Psi\fn{t,x} \Psi^\dagger\fn{s,x} \right>_\text{A-(iii)} \simeq \notag \\
	&
		+
		\frac{9 g_0^4}{\paren{4\pi}^4 {t}^2 m_0^2}
		\Bigg\{
		\frac{1}{\vep} + 1 + \log\sqbr{2m_0^2\fn{2s+t}} + \log\fn{ \frac{32\pi^2 \mu^4 {t}}{m_0^2} }
		+ e^{2m_0^2\fn{2s+t}} G^{2,0}_{1,2}\fn{ {}^{1}_{0,0} | \, 2 m_0^2 \fn{2s+t} }
		\Bigg\} + {\cal O}\fn{\vep}.
}
The form in Eq.~\eqref{eq:Psi-2PTform-A4p} has no UV divergence in $d \to 4$, but it contains an infrared divergence.
To regularize this, we deform the upper bound of the integral range of $y$ with a positive dimensionless parameter $a$ as follows:
\al{
\left< \Psi\fn{t,x} \Psi^\dagger\fn{s,x} \right>_\text{A-(vi)}
	&\simeq
		\frac{g_0^4}{\paren{4\pi}^4\paren{s+t}^2 m_0^2}
		\frac{1}{4}
		\int_0^1 \!\! \df x \int_0^{1-x} \!\!\!\! \df y
		\sqbr{ \frac{1}{ {1-x-y} } } \notag \\
	&\hookrightarrow
		\frac{g_0^4}{\paren{4\pi}^4\paren{s+t}^2 m_0^2}
		\frac{1}{4}
		\int_0^1 \!\! \df x \int_0^{1-x-a} \!\!\!\! \df y
		\sqbr{ \frac{1}{ {1-x-y} } } \notag \\
	&=
		\frac{g_0^4}{\paren{4\pi}^4\paren{s+t}^2 m_0^2}
		\frac{1}{4}
		\sqbr{-1-\log\fn{a}},
}
where the step designated by the symbol $\hookrightarrow$ corresponds to the regularization of an infrared divergence.

\subsubsection{Diagram B}

In Eq.~\eqref{eq:Psi-2PTform-B}, after taking the limit $y \to x$ safely,
we obtain the form
\al{
\left< \Psi\fn{t,x} \Psi^\dagger\fn{s,x} \right>_\text{B}
	&=
		\paren{- g_0^4 \mu^{4\vep}}
		\int_p \frac{e^{ -\paren{t+s}p^2}}{p^2 + m_0^2}
		\int_0^t \df u \int_\ell \frac{d}{\ell^2} e^{-2u \ell^2} \notag \\
	&=
		\paren{- g_0^4 \mu^{4\vep}}
		\int_p \frac{e^{ -\paren{t+s}p^2}}{p^2 + m_0^2}
		\int_0^t \df u  \paren{ \frac{2d}{\paren{4\pi}^{d/2}\Gamma\fn{d/2}} }
		\int_0^\infty \df\ell \ell^{d-3} e^{-2u \ell^2} \notag \\
	&=
		\paren{- g_0^4 \mu^{4\vep}}
		\int_p \frac{e^{ -\paren{t+s}p^2}}{p^2 + m_0^2}
		\int_0^t \df u  \paren{ \frac{2d}{\paren{4\pi}^{d/2}\Gamma\fn{d/2}} }
		2^{-d/2} u^{1-d/2} \Gamma\fn{-1 + \frac{d}{2}} \notag \\
	&=
		\paren{- g_0^4 \mu^{4\vep}}
		\int_p \frac{e^{ -\paren{t+s}p^2}}{p^2 + m_0^2}
		\frac{1}{\paren{4\pi}^{d/2}} \int_0^t \df u \frac{d}{ {d}/{2} - 1} \paren{2u}^{1-d/2} \notag \\
	&=
		\paren{- g_0^4 \mu^{4\vep}}
		\int_p \frac{e^{ -\paren{t+s}p^2}}{p^2 + m_0^2}
		\frac{1}{\paren{4\pi}^{2-\vep}} \frac{4-2\vep}{1-\vep} \frac{ \paren{2t}^\vep }{ 2\vep } \notag \\
	&\simeq
		\paren{- g_0^4 \mu^{4\vep}}
		\paren{ \frac{1}{4\pi\paren{s+t}} }^{d/2} \frac{1}{m_0^2}
		\frac{1}{\paren{4\pi}^{2-\vep}} \frac{4-2\vep}{1-\vep} \frac{ \paren{2t}^\vep }{ 2\vep },
		\label{eq:Psi-2PTform-Bp}
}
where in the second and third lines we integrated the isotropic angular part and the radial part of the $d$-dimensional $\ell$ integral, respectively, while in the last step we adopted the formula in Eq.~\eqref{eq:SPA-formula} to evaluate the $p$ integral around the saddle point $p_\ast = 0$.

It is very straightforward to obtain an $\vep$-expanded form of Eq.~\eqref{eq:Psi-2PTform-Bp}:
\al{
\left< \Psi\fn{t,x} \Psi^\dagger\fn{s,x} \right>_\text{B}
	&\simeq
		-\frac{g_0^4}{\paren{4\pi}^4\paren{s+t}^2 m_0^2}
		\br{
			\frac{2}{\vep} + 1 + 2 \log\sqbr{32\pi^2 t \paren{s+t} \mu^4}
		} + {\cal O}\fn{\vep}.
}

\subsubsection{Diagram C}

In Eq.~\eqref{eq:Psi-2PTform-C}, after taking the limit $y \to x$ safely,
we can perform the integrals on the flow time exactly, where Eq.~\eqref{eq:Psi-2PTform-C} leads to
\al{
\left< \Psi\fn{t,x} \Psi^\dagger\fn{s,x} \right>_\text{C}
	&=
		\paren{g_0^4 \mu^{4\vep}}
		\int_p \frac{e^{ -\paren{t+s}p^2}}{p^2 + m_0^2}
		\int_\ell \frac{ \paren{p+\ell}^2 }{ \paren{p-\ell}^2 \paren{\ell^2 + m_0^2} }
		\times
		\frac{1}{2 \paren{\ell^2 - \ell\cdot p}} \sqbr{ 1 - e^{-2t \paren{\ell^2 - \ell\cdot p}} },
}
where we divide this form into the following two pieces:
\al{
\left< \Psi\fn{t,x} \Psi^\dagger\fn{s,x} \right>_\text{C-(i)}
	&:=
		+
		\paren{g_0^4 \mu^{4\vep}}
		\int_p \frac{e^{ -\paren{t+s}p^2}}{p^2 + m_0^2}
		\int_\ell \frac{ \paren{p+\ell}^2 }{ \paren{p-\ell}^2 \paren{\ell^2 + m_0^2} }
		\times
		\frac{1}{2 \paren{\ell^2 - \ell\cdot p}}, 
		\label{eq:Psi-2PTform-C1} \\
\left< \Psi\fn{t,x} \Psi^\dagger\fn{s,x} \right>_\text{C-(ii)}
	&:=
		-
		\paren{g_0^4 \mu^{4\vep}}
		\int_p \frac{e^{ -\paren{t+s}p^2}}{p^2 + m_0^2}
		\int_\ell \frac{ \paren{p+\ell}^2 }{ \paren{p-\ell}^2 \paren{\ell^2 + m_0^2} }
		\times
		\frac{1}{2 \paren{\ell^2 - \ell\cdot p}} e^{-2t \paren{\ell^2 - \ell\cdot p}}.
		\label{eq:Psi-2PTform-C2}
}

For Eq.~\eqref{eq:Psi-2PTform-C1}, it is straightforward to perform the saddle-point integral of $p$ around the saddle point $p_\ast  = 0$ as
\al{
\left< \Psi\fn{t,x} \Psi^\dagger\fn{s,x} \right>_\text{C-(i)}
	&\simeq
		\frac{\paren{g_0^4 \mu^{4\vep}}}{2}
		\paren{ \frac{1}{4\pi\paren{s+t}} }^{d/2} \frac{1}{m_0^2}
		\int_\ell \frac{ 1 }{ \paren{\ell^2 + m_0^2} \ell^2 } \notag \\
	&=
		\frac{\paren{g_0^4 \mu^{4\vep}}}{2}
		\paren{ \frac{1}{4\pi\paren{s+t}} }^{d/2} \frac{1}{m_0^2}
		\frac{1}{\paren{4\pi}^{d/2}} \frac{ \Gamma\fn{2-d/2} }{\Gamma\fn{2}} \int_0^1 \df x \paren{ \frac{1}{x m_0^2} }^{2-d/2},
		\label{eq:Psi-2PTform-C1p}
}
where in the second step we integrate the $\ell$ part with the help of the Feynman parameterization in a standard method.

For Eq.~\eqref{eq:Psi-2PTform-C2}, we can perform the saddle-point integral of $\ell$ around the saddle point $\ell_\ast = p/2$ as
\al{
\left< \Psi\fn{t,x} \Psi^\dagger\fn{s,x} \right>_\text{C-(ii)}
	&\simeq
		\frac{\paren{9 \cdot 4^2 g_0^4 \mu^{4\vep}}}{2}
		\paren{ \frac{1}{8\pi{t}} }^{d/2} 
		\int_p \frac{1}{ p^2 \paren{p^2 + m_0^2} \paren{p^2 + 4 m_0^2} } e^{- \paren{ \frac{t}{2} + s } p^2 } \notag \\
	&=
		\frac{\paren{9 \cdot 4^2 g_0^4 \mu^{4\vep}}}{2}
		\paren{ \frac{1}{8\pi{t}} }^{d/2}
		\paren{ \frac{2}{ \paren{4\pi}^{d/2} \Gamma\fn{d/2} } } \notag \\
	&\quad
		\times
		\int_0^\infty \df p p^{d-1} \frac{1}{ p^2 \paren{p^2 + m_0^2} \paren{p^2 + 4 m_0^2} } e^{- \paren{ \frac{t}{2} + s } p^2 } \notag \\
	&=
		\frac{\paren{9 \cdot 4^2 g_0^4 \mu^{4\vep}}}{2}
		\paren{ \frac{1}{8\pi{t}} }^{d/2}
		\paren{ \frac{2}{ \paren{4\pi}^{d/2} \Gamma\fn{d/2} } } \notag \\
	&\quad
		\times
		\frac{1}{96 \Gamma\fn{2-d/2}} e^{ \frac{1}{2} \paren{ -id\pi + m_0^2\paren{2s+t} } }
		m_0^{-6+d} \pi \sqbr{ i + \cot\fn{\frac{d\pi}{2}} } \notag \\
	&\quad
		\times
		\sqbr{
			-16 \Gamma\fn{2-\frac{d}{2}, \frac{1}{2}m_0^2\paren{2s+t}}
			+2^d e^{3m_0^2\paren{2s+t}/2} \Gamma\fn{ 2-\frac{d}{2}, 2m_0^2\paren{2s+t} }
		},
		\label{eq:Psi-2PTform-C2p}
}
where in the second and third lines we integrated the isotropic angular part and the radial part of the $d$-dimensional $p$ integral, respectively.

The $\vep$ expansion of Eq.~\eqref{eq:Psi-2PTform-C1p} is easily obtained as
\al{
\left< \Psi\fn{t,x} \Psi^\dagger\fn{s,x} \right>_\text{C-(i)}
	&\simeq
		\frac{{g_0^4 }}{2}
		\paren{ \frac{1}{4\pi\paren{s+t}} }^{2} \frac{1}{m_0^2} \frac{1}{\paren{4\pi}^2}
		\sqbr{ \mu^{4\vep} \paren{4\pi\paren{t+s}}^{\vep} \paren{4\pi}^\vep } \Gamma\fn{\vep}
		\sqbr{ 1+\vep \paren{1 - \log{m_0^2}} } \notag \\
	&=
		\frac{g_0^4}{\paren{4\pi}^4\paren{s+t}^2 m_0^2}
		\frac{1}{2}
		\br{  \frac{1}{\vep} - \gamma + 1 + \log\sqbr{ \frac{16\pi^2\fn{s+t}\mu^4}{m_0^2} }  }
		+ {\cal O}\fn{\vep}.
}
Meanwhile, the following expansion forms,
\al{
\frac{1}{\Gamma\fn{\vep}\Gamma\fn{2-\vep}}
	&=
		\vep + {\cal O}\fn{\vep^2}, \\
e^{\pi i \vep}
	&=
		1 + i \pi \vep + {\cal O}\fn{\vep^2}, \\
\cot\sqbr{\pi \paren{2-\vep}}
	&=
		- \frac{1}{\pi \vep} + \frac{\pi}{3} \vep + {\cal O}\fn{\vep^2}, \\
\Gamma\fn{\vep, A}
	&=
		\Gamma\fn{0,A} + 
		\sqbr{
			\Gamma\fn{0,A} \log{A} + G^{2,0}_{1,2}\fn{ {}^{1}_{0,0} \Big| \, A }
		} \vep + {\cal O}\fn{\vep^2},
}
for a positive $A$, bring us to the final form of the $\vep$ expansion of Eq.~\eqref{eq:Psi-2PTform-C2p}:
\al{
&\left< \Psi\fn{t,x} \Psi^\dagger\fn{s,x} \right>_\text{C-(ii)} \notag \\
	&\simeq
		\frac{g_0^4}{\paren{4\pi}^4{t}^2 m_0^2}
		\frac{3}{8}
		e^{ \frac{1}{2} m_0^2 \paren{2s+t} }
		\br{ 16 \Gamma\fn{0, \frac{1}{2} m_0^2 \paren{2s+t}} - 16 e^{ \frac{3}{2}m_0^2\paren{2s+t} } \Gamma\fn{0, 2 m_0^2 \paren{2s+t}} }
		+ {\cal O}\fn{\vep}.
}

\subsubsection{Diagram D}

In Eq.~\eqref{eq:Psi-2PTform-D}, after taking the limit $y \to x$ safely,
we can perform the integrals on the flow times exactly, where Eq.~\eqref{eq:Psi-2PTform-D} leads to
\al{
\left< \Psi\fn{t,x} \Psi^\dagger\fn{s,x} \right>_\text{D}
	&=
		\paren{g_0^4 \mu^{4\vep}}
		\int_p \frac{e^{ -\paren{t+s}p^2}}{p^2 + m_0^2}
		\int_\ell \frac{ \paren{\ell+p}^2 }{ \paren{\ell-p}^2 } \notag \\
	&\quad
		\times
		\br{
			\frac{1}{4 \paren{\ell^2 - \ell\cdot p} \paren{p-\ell}^2} -
			\frac{e^{-2t\paren{\ell^2 - \ell\cdot p}}}{4 \paren{p^2 - p\cdot \ell} \paren{\ell^2 - \ell\cdot p}} -
			\frac{e^{-2t\paren{p-\ell}^2}}{4 \paren{p^2 - p\cdot \ell} \paren{p-\ell}^2}
		},
}
where we divide this form into the following three pieces:
\al{
\left< \Psi\fn{t,x} \Psi^\dagger\fn{s,x} \right>_\text{D-(i)}
	&:=
		+
		\paren{g_0^4 \mu^{4\vep}}
		\int_p \frac{e^{ -\paren{t+s}p^2}}{p^2 + m_0^2}
		\int_\ell \frac{ \paren{\ell+p}^2 }{ \paren{\ell-p}^2 }
		\times
		\frac{1}{4 \paren{\ell^2 - \ell\cdot p} \paren{p-\ell}^2},
		\label{eq:Psi-2PTform-D1} \\
\left< \Psi\fn{t,x} \Psi^\dagger\fn{s,x} \right>_\text{D-(ii)}
	&:=
		-
		\paren{g_0^4 \mu^{4\vep}}
		\int_p \frac{e^{ -\paren{t+s}p^2}}{p^2 + m_0^2}
		\int_\ell \frac{ \paren{\ell+p}^2 }{ \paren{\ell-p}^2 }
		\times
		\frac{e^{-2t\paren{\ell^2 - \ell\cdot p}}}{4 \paren{p^2 - p\cdot \ell} \paren{\ell^2 - \ell\cdot p}}, 
		\label{eq:Psi-2PTform-D2} \\
\left< \Psi\fn{t,x} \Psi^\dagger\fn{s,x} \right>_\text{D-(iii)}
	&:=
		-
		\paren{g_0^4 \mu^{4\vep}}
		\int_p \frac{e^{ -\paren{t+s}p^2}}{p^2 + m_0^2}
		\int_\ell \frac{ \paren{\ell+p}^2 }{ \paren{\ell-p}^2 }
		\times
		\frac{e^{-2t\paren{p-\ell}^2}}{4 \paren{p^2 - p\cdot \ell} \paren{p-\ell}^2}.
		\label{eq:Psi-2PTform-D3}
}

For Eq.~\eqref{eq:Psi-2PTform-D1}, we evaluate the $p$ integral using Eq.~\eqref{eq:SPA-formula} around the saddle point $p_\ast = 0$ as
\al{
\left< \Psi\fn{t,x} \Psi^\dagger\fn{s,x} \right>_\text{D-(i)}
	&\simeq
		\paren{g_0^4 \mu^{4\vep}}
		\paren{ \frac{1}{4\pi\paren{s+t}} }^{d/2}
		\frac{1}{4m_0^2}
		\int_\ell \frac{1}{\paren{\ell^2}^2} \notag \\
	&\hookrightarrow
		\paren{g_0^4 \mu^{4\vep}}
		\paren{ \frac{1}{4\pi\paren{s+t}} }^{d/2}
		\frac{1}{4m_0^2}
		\int_\ell \frac{1}{\paren{\ell^2 + \mu_\text{IR}^2}^2} \notag \\
	&=
		\paren{g_0^4 \mu^{4\vep}}
		\paren{ \frac{1}{4\pi\paren{s+t}} }^{d/2}
		\frac{1}{4m_0^2}
		\frac{1}{\paren{4\pi}^{d/2}} \frac{\Gamma\fn{2-d/2}}{\Gamma\fn{2}} \paren{\frac{1}{\mu_\text{IR}^2}}^{2-d/2},
		\label{eq:Psi-2PTform-D1p}
}
where in the second line we introduced a virtual mass $\mu_\text{IR}$ to regularize the integral.
The calculation in the third line was performed in a standard technique.

For Eq.~\eqref{eq:Psi-2PTform-D2}, the $\ell$ integral can be performed in the saddle point method in Eq.~\eqref{eq:SPA-formula}
around the saddle point $\ell_\ast = p/2$ as
\al{
\left< \Psi\fn{t,x} \Psi^\dagger\fn{s,x} \right>_\text{D-(ii)}
	&\simeq
		18 \paren{g_0^4 \mu^{4\vep}}
		\paren{ \frac{1}{8\pi t} }^{d/2}
		\int_p \frac{1}{\paren{p^2}^2 \paren{p^2 + m_0^2}} e^{- \paren{s+t/2}p^2} \notag \\
	&=
		18 \paren{g_0^4 \mu^{4\vep}}
		\paren{ \frac{1}{8\pi t} }^{d/2}
		\paren{ \frac{2}{ \paren{4\pi}^{d/2} \Gamma\fn{d/2} } }
		\int_0^\infty \df p \, p^{d-1} \frac{1}{\paren{p^2}^2 \paren{p^2 + m_0^2}} e^{- \paren{s+t/2}p^2} \notag \\
	&=
		18 \paren{g_0^4 \mu^{4\vep}}
		\paren{ \frac{1}{8\pi t} }^{d/2}
		\paren{ \frac{2}{ \paren{4\pi}^{d/2} \Gamma\fn{d/2} } } \notag \\
	&\quad
		\times
		\frac{1}{2} e^{m_0^2\paren{2s+t}/2} m_0^{-6+d}
		\Gamma\fn{-2+\frac{d}{2}} \Gamma\fn{3-\frac{d}{2}, \frac{1}{2} m_0^2 \paren{2s+t}},
		\label{eq:Psi-2PTform-D2p}
}
where in the second and third lines we integrated the isotropic angular part and the radial part of the $d$-dimensional $p$ integral, respectively.

For Eq.~\eqref{eq:Psi-2PTform-D3}, the exponent is square-completed as
\al{
-\paren{s+t} p^2 -2t \paren{p-\ell}^2
	&=
		-\paren{3t+s} \sqbr{ p - \frac{2t}{3t+s} \ell }^2 - \frac{2t\paren{t+s}}{3t+s} \ell^2,
}
which tells us the saddle point of the $p$ integral as $p_\ast = \sqbr{2t/\paren{3t+s}}\ell$.
Through the formula in Eq.~\eqref{eq:SPA-formula}, we get
\al{
\left< \Psi\fn{t,x} \Psi^\dagger\fn{s,x} \right>_\text{D-(iii)}
	&\simeq
		\frac{\paren{g_0^4 \mu^{4\vep}}}{4}
		\paren{ \frac{1}{4\pi\paren{3t+s}} }^{d/2}
		\frac{ \paren{1+\frac{2t}{3t+s}}^2 \paren{- \frac{\paren{3t+s}^2}{2t\paren{t+s}} } }{ \paren{1-\frac{2t}{3t+s}}^4 \paren{\frac{2t}{3t+s}}^2 }
		\notag \\
	&\quad
		\times
		\int_\ell \frac{1}{\paren{\ell^2}^2} \frac{1}{ \ell^2 + \paren{\frac{3t+s}{2t}}^2 m_0^2 }
		e^{- \frac{2t\paren{t+s}}{3t+s} \ell^2}.
}
The structure of the above $\ell$ integral is the same as that of Diagram D-(ii), and we immediately obtain the final form:
\al{
\left< \Psi\fn{t,x} \Psi^\dagger\fn{s,x} \right>_\text{D-(iii)}
	&\simeq
		\frac{\paren{g_0^4 \mu^{4\vep}}}{4}
		\paren{ \frac{1}{4\pi\paren{3t+s}} }^{d/2}
		{\cal A}
		\notag \\
	&\quad
		\times
		\paren{ \frac{2}{ \paren{4\pi}^{d/2} \Gamma\fn{d/2} } } 
		\frac{1}{2} e^{{\cal C} {\cal M}^2} {\cal M}^{-6+d}
		\Gamma\fn{-2+\frac{d}{2}} \Gamma\fn{3-\frac{d}{2}, {\cal C} {\cal M}^2 },
		\label{eq:Psi-2PTform-D3p}
}
where we define the variables for our convenience,
\al{
{\cal A}
	&=
		- \frac{ \paren{s+3t}^4 }{ \paren{2t}^3 \paren{s+t}}
		\frac{ \paren{1 + \frac{2t}{s+3t}}^2 }{ \paren{1 - \frac{2t}{s+3t}}^4 },&
{\cal M}
	&=
		\paren{ \frac{s+3t}{2t} } m_0,&
{\cal C}
	&=
		\frac{ 2t \paren{s+t} }{ s+3t }.&
}

The $\vep$-expanded form of Eq.~\eqref{eq:Psi-2PTform-D1p} is evaluated straightforwardly as
\al{
\left< \Psi\fn{t,x} \Psi^\dagger\fn{s,x} \right>_\text{D-(i)}
	&\simeq
		\frac{g_0^4}{\paren{4\pi}^4\paren{s+t}^2 m_0^2}
		\frac{1}{4}
		\br{ \frac{1}{\vep} - \gamma + \log\sqbr{  \frac{16\pi^2\fn{s+t}\mu^4}{\mu_\text{IR}^2}  } } + {\cal O}\fn{\vep}.
}
Calculations of the forms in Eqs.~\eqref{eq:Psi-2PTform-D2p} and \eqref{eq:Psi-2PTform-D3p} are also straightforward with the help of
the formulas in Eqs.~\eqref{eq:vep-formula-1} and \eqref{eq:vep-formula-2} as
\al{
&\left< \Psi\fn{t,x} \Psi^\dagger\fn{s,x} \right>_\text{D-(ii)} \notag \\
	&\simeq
		-
		\frac{g_0^4}{\paren{4\pi}^4{t}^2 m_0^2}
		\frac{9}{2}
		\br{
			 \frac{1}{\vep} + 1 + \log\fn{\frac{1}{2}m_0^2\fn{2s+t}} + \log\fn{ \frac{32\pi^2t\mu^4}{m_0^2}  } 
			+ e^{\frac{1}{2}m_0^2\fn{2s+t}} G^{2,0}_{1,2}\fn{ {}^{1}_{0,0} \Big| \, \frac{1}{2} m_0^2 \fn{2s+t} }
		} + {\cal O}\fn{\vep}, \\
&\left< \Psi\fn{t,x} \Psi^\dagger\fn{s,x} \right>_\text{D-(iii)} \notag \\
	&\simeq 
		-
		\frac{g_0^4}{\paren{4\pi}^4\paren{s+3t}^2 {\cal M}^2}
		\frac{ {\cal A} }{4}
		\br{
			 \frac{1}{\vep} + 1 + \log\fn{ {\cal C} {\cal M}^2 } + \log\sqbr{ \frac{16\pi^2\fn{s+3t}\mu^4}{{\cal M}^2}  } 
			+ e^{ {\cal C} {\cal M}^2 } G^{2,0}_{1,2}\fn{ {}^{1}_{0,0} \Big| \, {\cal C} {\cal M}^2 } 
		} + {\cal O}\fn{\vep}. 
}

\subsubsection{Diagram E}

After taking the safe limit $y \to x$ as in Eq.~\eqref{eq:Psi-2PTform-E}, we obtain
\al{
\left< \Psi\fn{t,x} \Psi^\dagger\fn{s,x} \right>_\text{E}
	&=
		\paren{-g_0^4 \mu^{4\vep}}
		\int_p \frac{e^{-\paren{t+s}p^2}}{p^2 + m_0^2}
		\int_0^t \df u \int_0^u \df\wt{u} \int_\ell \frac{ \paren{\ell+p}^2 }{ \ell^2 + m_0^2 }
		e^{ -2u \ell^2 + 2 \paren{u-\wt{u}} \ell\cdot p }.
		\label{eq:Psi-2PTform-E-I}
%
}

We can perform the integrals on the flow times exactly, where Eq.~\eqref{eq:Psi-2PTform-E-I} leads to
\al{
\left< \Psi\fn{t,x} \Psi^\dagger\fn{s,x} \right>_\text{E}
	&=
		\paren{-g_0^4 \mu^{4\vep}}
		\int_p \frac{e^{-\paren{t+s}p^2}}{p^2 + m_0^2}
		\int_\ell \frac{ \paren{\ell+p}^2 }{ \ell^2 + m_0^2 } \notag \\
	&\quad
		\times
		\sqbr{
			\frac{1}{4 \ell^2 \paren{\ell^2 - \ell\cdot p}} -
			\frac{1}{4 \paren{\ell\cdot p} \paren{\ell^2 - \ell\cdot p}} e^{-2t \paren{\ell^2 - \ell\cdot p}} +
			\frac{1}{4 \paren{\ell \cdot p} \ell^2} e^{-2t \ell^2}
		},
}
where we divide this form into the following three pieces:
\al{
\left< \Psi\fn{t,x} \Psi^\dagger\fn{s,x} \right>_\text{E-(i)}
	&:=
		-
		\paren{g_0^4 \mu^{4\vep}}
		\int_p \frac{e^{-\paren{t+s}p^2}}{p^2 + m_0^2}
		\int_\ell \frac{ \paren{\ell+p}^2 }{ \ell^2 + m_0^2 }
		\times
		\frac{1}{4 \ell^2 \paren{\ell^2 - \ell\cdot p}},
		\label{eq:Psi-2PTform-E-I1} \\
\left< \Psi\fn{t,x} \Psi^\dagger\fn{s,x} \right>_\text{E-(ii)}
	&:=
		+
		\paren{g_0^4 \mu^{4\vep}}
		\int_p \frac{e^{-\paren{t+s}p^2}}{p^2 + m_0^2}
		\int_\ell \frac{ \paren{\ell+p}^2 }{ \ell^2 + m_0^2 }
		\times
		\frac{1}{4 \paren{\ell\cdot p} \paren{\ell^2 - \ell\cdot p}} e^{-2t \paren{\ell^2 - \ell\cdot p}},
		\label{eq:Psi-2PTform-E-I2} \\
\left< \Psi\fn{t,x} \Psi^\dagger\fn{s,x} \right>_\text{E-(iii)}
	&:=
		-
		\paren{g_0^4 \mu^{4\vep}}
		\int_p \frac{e^{-\paren{t+s}p^2}}{p^2 + m_0^2}
		\int_\ell \frac{ \paren{\ell+p}^2 }{ \ell^2 + m_0^2 }
		\times
		\frac{1}{4 \paren{\ell \cdot p} \ell^2} e^{-2t \ell^2}.
		\label{eq:Psi-2PTform-E-I3}
}

For Eq.~\eqref{eq:Psi-2PTform-E-I1}, we easily find the saddle point $p_\ast = 0$, and utilizing the formula in Eq.~\eqref{eq:SPA-formula}
leads to
\al{
\left< \Psi\fn{t,x} \Psi^\dagger\fn{s,x} \right>_\text{E-(i)}
	&\simeq
		-
		\frac{\paren{g_0^4 \mu^{4\vep}}}{4}
		\paren{\frac{1}{4\pi\paren{t+s}}}^{d/2} \frac{1}{m_0^2}
		\int_\ell \frac{1}{\ell^2 \paren{\ell^2 + m_0^2}} \notag \\
	&=
		-
		\frac{\paren{g_0^4 \mu^{4\vep}}}{4}
		\paren{\frac{1}{4\pi\paren{t+s}}}^{d/2} \frac{1}{m_0^2}
		\frac{1}{\paren{4\pi}^{d/2}} \frac{\Gamma\fn{2-d/2}}{\Gamma\fn{2}} \paren{\frac{1}{x m_0^2}}^{2-d/2},
		\label{eq:Psi-2PTform-E-I1p}
}
where the second step is an ordinary calculation with the help of the Feynman parameterization.

For Eq.~\eqref{eq:Psi-2PTform-E-I2}, the saddle point of $\ell$ is located at $\ell_\ast = p/2$, and the formula in Eq.~\eqref{eq:SPA-formula} brings us to
\al{
\left< \Psi\fn{t,x} \Psi^\dagger\fn{s,x} \right>_\text{E-(ii)}
	&\simeq
		-
		\paren{9 \cdot 2 g_0^4 \mu^{4\vep}}
		\paren{\frac{1}{8\pi t}}^{d/2} 
		\int_p \frac{1}{p^2 \paren{p^2 + m_0^2} \paren{p^2 + 4m_0^2}} e^{-\paren{s+t/2}p^2} \notag \\
	&=
		-
		\paren{18 g_0^4 \mu^{4\vep}}
		\paren{\frac{1}{8\pi t}}^{d/2}
		\paren{ \frac{2}{ \paren{4\pi}^{d/2} \Gamma\fn{d/2} } } \notag \\
	&\quad
		\times
		\int_0^\infty \df p p^{d-1} \frac{1}{ p^2 \paren{p^2 + m_0^2} \paren{p^2 + 4 m_0^2} } e^{- \paren{ \frac{t}{2} + s } p^2 } \notag \\
	&=
		-
		\paren{18 g_0^4 \mu^{4\vep}}
		\paren{\frac{1}{8\pi t}}^{d/2}
		\paren{ \frac{2}{ \paren{4\pi}^{d/2} \Gamma\fn{d/2} } } \notag \\
	&\quad
		\times
		\frac{1}{96 \Gamma\fn{2-d/2}} e^{ \frac{1}{2} \paren{ -id\pi + m_0^2\paren{2s+t} } }
		m_0^{-6+d} \pi \sqbr{ i + \cot\fn{\frac{d\pi}{2}} } \notag \\
	&\quad
		\times
		\sqbr{
			-16 \Gamma\fn{2-\frac{d}{2}, \frac{1}{2}m_0^2\paren{2s+t}}
			+2^d e^{3m_0^2\paren{2s+t}/2} \Gamma\fn{ 2-\frac{d}{2}, 2m_0^2\paren{2s+t} }
		},
		\label{eq:Psi-2PTform-E-I2p}
}
where in the second and third lines we integrated the isotropic angular part and the radial part of the $d$-dimensional $p$ integral, respectively.

For Eq.~\eqref{eq:Psi-2PTform-E-I3}, we can make a deformation with a Feynman parameter integral,
\al{
\left< \Psi\fn{t,x} \Psi^\dagger\fn{s,x} \right>_\text{E-(iii)}
	&=
		-
		\frac{\paren{g_0^4 \mu^{4\vep}}}{4}
		\int_{p,\ell} \frac{\paren{p+\ell}^2}{p^2+m_0^2}
		\int_0^1 \!\! \df x \int_0^{1-x} \!\! \df y 
		\frac{2! e^{-\paren{t+s}p^2 - 2t \ell^2 }}
		{ \sqbr{ \paren{x+y} \ell^2 + y m_0^2 + \paren{1-x-y}\paren{p\cdot\ell} }^3 } \notag \\
	&\simeq
		-
		\frac{\paren{g_0^4 \mu^{4\vep}}}{4}
		\int_\ell \frac{\ell^2}{m_0^2} \int_0^1 \df x \int_0^{1-x} \df y 
		\frac{2! e^{ - 2t \ell^2 }}
		{ \sqbr{ \paren{x+y} \ell^2 + y m_0^2 }^3 } \notag \\
	&\simeq
		0,
}
where in the second and the third steps we performed the $p$ and $\ell$ integrals using the formula in Eq.~\eqref{eq:SPA-formula}
around the saddle points $p_\ast = 0$ and $\ell_\ast = 0$, respectively.


The $\vep$ expansion of Eq.~\eqref{eq:Psi-2PTform-E-I1p} is straightforward, and that of Eq.~\eqref{eq:Psi-2PTform-E-I2p} is doable following the process for Eq.~\eqref{eq:Psi-2PTform-C2p}, where their results are
\al{
\left< \Psi\fn{t,x} \Psi^\dagger\fn{s,x} \right>_\text{E-(i)}
	&\simeq
		- \frac{g_0^4}{\paren{4\pi}^4\paren{s+t}^2 m_0^2}
		\frac{1}{4}
		\br{ \frac{1}{\vep} - \gamma +1 + \log\sqbr{  \frac{16\pi^2\paren{s+t}\mu^4}{m_0^2}  } }
		+ {\cal O}\fn{\vep}, \\
\left< \Psi\fn{t,x} \Psi^\dagger\fn{s,x} \right>_\text{E-(ii)}
	&\simeq
		-
		\frac{g_0^4}{\paren{4\pi}^4{t}^2 m_0^2}
		\frac{3}{32}
		e^{ \frac{1}{2} m_0^2 \paren{2s+t} } \notag \\
	&\quad
		\times
		\sqbr{ 16 \Gamma\fn{0, \frac{1}{2} m_0^2 \paren{2s+t}}
			 - 16 \, e^{ \frac{3}{2}m_0^2\paren{2s+t} } \Gamma\fn{0, 2 m_0^2 \paren{2s+t}} }
		+ {\cal O}\fn{\vep}.
}

\subsubsection{Diagram F} 

{
We can perform the integrals on the flow times exactly, where Eq.~\eqref{eq:Psi-2PTform-F} leads to
\al{
\left< \Psi\fn{t,x} \Psi^\dagger\fn{s,x} \right>_\text{F}
	&=
		\paren{-4 g_0^4 \mu^{4\vep}}
		\int_{p,\ell} \frac{e^{ -\paren{t+s}p^2}}{p^2 + m_0^2}
		\frac{ p_\mu \ell_\mu }{ \ell^2 + m_0^2 }
		\times
		\sqbr{
			\frac{2 \ell^2 t - 1}{4 \paren{\ell^2}^2} + \frac{1}{4 \paren{\ell^2}^2} e^{-2t \ell^2}
		},
}
where this integral is odd for $p_\mu$ and $\ell_\mu$, and the result is zero.}


\subsubsection{Diagram a}

For Eq.~\eqref{eq:Psi-2PTform-a}, after taking the limit $y \to x$ safely,
we find that the $p$ integral part and $\ell$ integral part are separable,
\al{
\left< \Psi\fn{t,x} \Psi^\dagger\fn{s,x} \right>_\text{a}
	&=
\paren{-d g_0^4 \mu^{4\vep}}
		\int_p \frac{e^{-\paren{t+s}p^2}}{ \paren{p^2 + m_0^2}^2 }
		\int_\ell \frac{1}{ \ell^2 + \mu_A^2 },
}
where evaluating the latter part is straightforwardly doable,
\al{
\int_\ell \frac{1}{ \ell^2 + \mu_A^2 }
	&=
		\frac{1}{\paren{4\pi}^{d/2}} \paren{ \mu_A^2 }^{1-\vep} \Gamma\fn{\vep-1},
		\label{eq:Diagram-a-ell-part}
}
which contains no infrared divergence, and we can take the limit $\mu_A \to 0$ safely.
Thereby, we conclude that
\al{
\left< \Psi\fn{t,x} \Psi^\dagger\fn{s,x} \right>_\text{a}
	&\xrightarrow[{\mu_A \to 0}]{}
		0.
}

\subsubsection{Diagram b}

For Eq.~\eqref{eq:Psi-2PTform-b}, after taking the limit $y \to x$ safely,
we can perform the $p$ integral part and $\ell$ integral part separately,
\al{
\left< \Psi\fn{t,x} \Psi^\dagger\fn{s,x} \right>_\text{b}
	&=
		\paren{- 4 \lambda_0 g_0^2 \mu^{4\vep}}
		\int_p \frac{e^{ -\paren{t+s}p^2}}{ \paren{p^2 + m_0^2}^2 }
		\int_\ell \frac{1}{ \ell^2 + m_0^2 },
}
where the latter part can be evaluated as in Eq.~\eqref{eq:Diagram-a-ell-part}.
The former integral is also analytically feasible
with the help of the exponential integral function $E_n\fn{z}$ as
\al{
\int_p \frac{e^{ -\paren{t+s}p^2}}{ \paren{p^2 + m_0^2}^2 }
	&=
		\paren{ \frac{2}{\paren{4\pi}^{d/2} \Gamma\fn{d/2}} }	
		\frac{1}{8} \paren{d-4} \paren{s+t}^{2-d/2} \Gamma\fn{-2+ \frac{d}{2} } \notag \\
	&\quad
		\times
		\br{
			-2 + e^{m_0^2 \paren{s+t}} \sqbr{-2+d+2m_0^2\paren{s+t}}
				E_{\frac{d-2}{2}}\fn{m_0^2\paren{s+t}}
		}.
} 
The total form is given as
\al{
\left< \Psi\fn{t,x} \Psi^\dagger\fn{s,x} \right>_\text{b}
	&=
		\paren{- 4 \lambda_0 g_0^2 \mu^{4\vep}} \frac{\paren{m_0^2}^{1-\vep}}{\paren{4\pi}^{4-2\vep}} 
		\Gamma\fn{\vep-1} \Gamma\fn{-\vep} \frac{2}{\Gamma\fn{2-\vep}} \paren{- \frac{\vep}{4}} \paren{s+t}^\vep \notag \\
	&\quad
		\times
		\br{
			-2 + e^{m_0^2 \paren{s+t}} \sqbr{2-2\vep + 2m_0^2\paren{s+t}}
				E_{1-\vep}\fn{m_0^2\paren{s+t}}
		}.
}

With the help of the following formulas,
\al{
\frac{\Gamma\fn{\vep-1} \Gamma\fn{-\vep}}{\Gamma\fn{2-\vep}}
	&=
		\frac{1}{\vep^2} + \frac{2-\gamma}{\vep} + \frac{1}{12} \paren{36-24\gamma+6\gamma^2 + \pi^2} + {\cal O}\fn{\vep}, \\
E_{1-\vep}\fn{m_0^2\paren{s+t}}
	&=
		E_{1}\fn{m_0^2\paren{s+t}} + G^{3,0}_{2,3}\fn{ {}^{1,1}_{0,0,0} \Big| \, m_0^2\paren{s+t} } \vep + {\cal O}\fn{\vep},
}
we reach the $\vep$-expanded form,
\al{
\left< \Psi\fn{t,x} \Psi^\dagger\fn{s,x} \right>_\text{b}
	&=
4 \lambda_0 g_0^2 \frac{m_0^2}{\paren{4\pi}^4}
		\Bigg\{
			\frac{1}{\vep} \sqbr{  -1 + e^{m_0^2\paren{s+t}} \paren{1 + m_0^2\paren{s+t}} E_1\fn{ m_0^2\paren{s+t} } } \notag \\
	&\quad \quad +
		\paren{ 2-\gamma + \log\sqbr{\frac{16\pi^2\paren{s+t}\mu^4}{m_0^2}} } 
			\sqbr{  -1 + e^{m_0^2\paren{s+t}} \paren{1 + m_0^2\paren{s+t}} E_1\fn{ m_0^2\paren{s+t} } } \notag \\
	&\quad \quad -
		e^{m_0^2\paren{s+t}} E_1\fn{ m_0^2\paren{s+t} } +
		e^{m_0^2\paren{s+t}} \paren{1 + m_0^2\paren{s+t}}
		G^{3,0}_{2,3}\fn{ {}^{1,1}_{0,0,0} \Big| \, m_0^2\paren{s+t} }
	\Bigg\} + {\cal O}\fn{\vep}.
}

\subsubsection{Diagram c}

For Eq.~\eqref{eq:Psi-2PTform-c}, the form after taking the limit $y \to x$ safely is
\al{
\left< \Psi\fn{t,x} \Psi^\dagger\fn{s,x} \right>_\text{c}
	&=
		{\paren{ g_0^4 \mu^{4\vep}}}
		\int_p \frac{e^{-\paren{t+s}p^2}}{ \paren{p^2 + m_0^2}^2 }
		\int_\ell \frac{ \paren{\ell+p}^2 }{  \paren{\ell-p}^2 \paren{\ell^2+m_0^2}  } \notag \\
	&\simeq
		{\paren{ g_0^4 \mu^{4\vep}}}
		\frac{1}{m_0^4} \paren{\frac{1}{4\pi\paren{s+t}}}^{d/2}
		\int_\ell \frac{ 1 }{  \paren{\ell^2+m_0^2}  } \notag \\
	&=
		{\paren{ g_0^4 \mu^{4\vep}}}
		\frac{1}{m_0^4} \paren{\frac{1}{4\pi\paren{s+t}}}^{d/2}
		\frac{1}{\paren{4\pi}^{d/2}} \paren{m_0^2}^{1-\vep} \Gamma\fn{\vep-1},
}
where in the second line we performed the saddle-point integration 
using Eq.~\eqref{eq:SPA-formula} with $p_\ast  = 0$, and in the third line,
Eq.~\eqref{eq:Diagram-a-ell-part} was applied.

The evaluation of the $\vep$-expanded form is straightforward, where the resultant form is
\al{
\left< \Psi\fn{t,x} \Psi^\dagger\fn{s,x} \right>_\text{c}
	&\simeq
		\frac{g_0^4}{\paren{4\pi}^4\paren{s+t}^2 m_0^2}
		\br{  -\frac{1}{\vep} + \gamma -1 + \log\sqbr{  \frac{m_0^2}{16\pi^2\paren{s+t}\mu^4}  }  } + {\cal O}\fn{\vep}.
}

\bibliographystyle{JHEP}
\bibliography{refs}

\end{document}